\pdfoutput=1
\documentclass[iop]{emulateapj}
\slugcomment{}

\shorttitle{Cluster-forming site AFGL 5157: colliding filamentary clouds and star formation}
\shortauthors{L.~K. Dewangan}
\begin{document}

\title{Cluster-forming site AFGL 5157: colliding filamentary clouds and star formation}
\author{L.~K. Dewangan\altaffilmark{1}}
\email{lokeshd@prl.res.in}
\altaffiltext{1}{Physical Research Laboratory, Navrangpura, Ahmedabad - 380 009, India.}
\begin{abstract}
We observationally investigate star formation (SF) process occurring in AFGL 5157 (area $\sim$13.5 pc $\times$ 13.5 pc) 
using a multi-wavelength approach. Embedded filaments are seen in the {\it Herschel} column density map, and one of them is 
identified as an elongated filamentary feature (FF) (length $\sim$8.3 pc; mass $\sim$1170 M$_{\odot}$).
Five {\it Herschel} clumps (M$_{clump}$ $\sim$45--300 M$_{\odot}$) are traced in the central part of FF, 
where an extended temperature structure (T$_{d}$ $\sim$13.5--26.5~K) is observed. 
In the direction of the central part of FF, the warmer region at T$_{d}$ $\sim$20--26.5~K spatially coincides with a mid-infrared (MIR) shell 
surrounding a previously known evolved infrared cluster. 
Diffuse H$\alpha$ emission is traced inside the infrared shell, suggesting the presence of massive stars in the evolved cluster. 
Based on the surface density analysis of young stellar objects (YSOs), embedded clusters of YSOs are traced toward the central part of FF, 
and are distributed around the infrared shell. Previously detected H$_{2}$O masers, H$_{2}$ knots, massive protostar candidates, and H\,{\sc ii} region are also seen toward the embedded clusters. 
Using the $^{12}$CO and $^{13}$CO line data, the central part of FF is observed at the overlapping zones of two filamentary molecular 
clouds (length $\sim$12.5 pc) around $-$20 and $-$17 km s$^{-1}$, which are also connected in velocity. 
Our observational results suggest that the formation of massive stars appears to be 
triggered by a collision of two filamentary molecular clouds, which might have also influenced the birth of YSOs in AFGL 5157. 
\end{abstract}
\section{Introduction}
\label{sec:intro}
In the era of the space based telescopes like {\it Spitzer} and {\it Herschel}, the observational detections of mid-infrared bubbles/shells 
and filaments have provided several crucial inputs to researchers for understanding the star formation (SF)
processes in Galactic star-forming sites \citep[e.g.,][]{churchwell06,andre10,andre14}. 
These observational inputs have been obtained through a detailed analysis of the infrared and sub-millimeter (sub-mm) data. 
Currently, the study concerning the involvement of filaments in the formation process of dense massive star-forming clumps and young stellar clusters 
has been received significant attention in the field of SF, which is still debated. 
In particular, the zones of merging/collision/interaction of filaments are considered a suitable environment for formation of massive OB stars ($\geq$ 8 M$_{\odot}$) and stellar clusters, where one may obtain higher column densities \citep[e.g.,][]{myers09,andre10,andre14,galvan10,duarte11,schneider12,inoue13,nakamura12,nakamura14,henshaw13,fukui14,dewangan17,dewangan19a,dewangan19}.
In the hydrodynamical simulations of the collision process, massive clumps and cores are 
produced at the junction of two filamentary molecular clouds or the shock-compressed interface layer \citep[e.g.,][and references therein]{habe92,anathpindika10,inoue13,haworth15a,haworth15b,torii17,balfour17,bisbas17}, where massive stars can be formed. However, the observational confirmation of such physical process or a colliding event is a very difficult task, and can be investigated via the knowledge of molecular gas motion toward the filaments and the zones of interacting/colliding filaments.

Situated at a distance of 1.8 kpc \citep{snell88,torrelles92a}, AFGL 5157/IRAS 05345+3157 is an active site of SF, which is also known as NGC 1985 or Mol 11. In the direction of IRAS 05345+3157, an extended shell detected in H$_{2}$ was reported, which was found to surround a cluster of infrared sources centered on the IRAS position \citep{chen99,chen03,reipurth08,varricatt10,wolf17}. 
\citet{chen99} suggested the presumptive age of this infrared cluster to be 10$^{6}$ yr, which was referred to as an evolved 
cluster in AFGL 5157 \citep[see also][]{chen03}. These authors suggested that massive stars located in this evolved cluster can be responsible for the H$_{2}$ shell through their UV fluorescence. They also reported the H$_{2}$ shell as an expanding shell. 
Using the {\it Spitzer} 3.6--8.0 $\mu$m images, the infrared emission has been 
observed toward the H$_{2}$ shell \citep{jiang13}, which is now referred to as an infrared shell in the present work.

Dust continuum, CS, and NH$_{3}$ emissions were observed about 1$'$.5 away from the evolved cluster or the infrared shell \citep{verdes89,pastor91,torrelles92b,klein05,lee11,fontani12,jiang13}. 
A few H$_{2}$O masers and an H\,{\sc ii} region were reported toward the dense NH$_{3}$ core \citep{torrelles92b}. Using the VLA 3.6 cm map, \citet{torrelles92b} suggested that the H\,{\sc ii} region is powered by a ZAMS B3 star. \citet{snell88} reported a bipolar CO outflow, which was centered on the dense NH$_{3}$ core \citep[see also][]{zhang05}. 
On the basis of previously published results in AFGL 5157, a summary figure containing different observed features (i.e., infrared shell, dense core, and H\,{\sc ii} region) is presented in Figure~\ref{fig1}a. 
Figure~\ref{fig1}a shows the {\it Spitzer} 3.6 $\mu$m image (size $\sim$2.7 pc $\times$ 2.7 pc) overlaid with the SCUBA 850 $\mu$m dust continuum emission contours \citep{difrancesco08,jiang13}. Three peaks seen in the 850 $\mu$m dust continuum map (i.e., peak1, peak2, and peak3) are also highlighted in the figure, which may trace the dense regions in AFGL 5157 \citep[see also Figure~18 in][]{reipurth08}.   
The infrared shell is also highlighted by the 3.6 $\mu$m continuum emission contour. 
The locations of seven previously observed CS cores \citep{lee11} are also marked in Figure~\ref{fig1}a, and are found toward 
two sub-mm peaks (i.e., peak1 and peak2). Note that the sub-mm peak3 was not covered in the CS observations \citep[see][]{lee11}.   
The direction in the figure is shown in the Galactic coordinates.

Previously, the signatures of ongoing SF were reported through the detection of molecular outflow, H$_{2}$O masers, H\,{\sc ii} region, and several H$_{2}$ knots distributed toward 
the dense NH$_{3}$ core \citep[see Figure~6 in][]{wolf17}. All these signposts are seen toward the sub-mm 850 $\mu$m continuum peak1 (see Figure~\ref{fig1}a). A detailed investigation of the sub-mm 850 $\mu$m continuum source (i.e., peak3) is not available in the literature. 
Recently, using high-resolution NH$_{3}$ line observations (resolution $\sim$2$''$.5), \citet{fontani12} observed two 
ammonia filaments in the direction of this dense NH$_{3}$ core/region (see Figure~1 in their paper). 
These authors traced the dense regions with NH$_{3}$ (1, 1) emission in a velocity range of [$-$17, $-$18.4] km s$^{-1}$ (see Table~1 in their paper).  
They found the seven CS cores, observed by \citet{lee11}, were within the ammonia filaments.  
Based on the observed results toward the dense NH$_{3}$ core and the H$_{2}$ shell, at least two distinct 
epochs of SF were reported in AFGL 5157 \citep[e.g.,][]{wolf17}. 
The evolved SF phase has been reported toward the infrared cluster or the H$_{2}$ shell, while the younger phase of 
SF (timescale $\sim$10$^{4}$--10$^{5}$ yr) is thought to be associated with the dense NH$_{3}$ core \citep[see also Figure~3 in][]{chen03}. 
It was also argued that the evolved infrared cluster might have triggered the younger phase of SF associated with 
the dense NH$_{3}$ core and northwest areas \citep{chen03}. 
Note that all the earlier works were mainly focused to an area of $\sim$1.8 pc $\times$ $\sim$1.8 pc containing AFGL 5157. 

The study of the physical environment at large-scale ($>$ 10 pc) is not yet reported in AFGL 5157. 
We do not find any study for examining embedded filaments, young stellar objects (YSOs), and 
the velocity structure of molecular gas on a wide-scale around AFGL 5157. 
Our present paper focuses on understanding the exact SF mechanisms occurring in AFGL 5157. 
In this context, a multi-wavelength approach has been adopted in this paper. A detailed 
analysis of $^{12}$CO(1-0) and $^{13}$CO(1-0) gas has been 
performed comparing its distribution against the infrared and sub-mm images. 
To trace the signatures of SF activity, a careful analysis of embedded YSOs is 
carried out for a large-scale area containing AFGL 5157.  

In Section~\ref{sec:obser}, we present details of the adopted data sets. 
Section~\ref{sec:data} provides the information of the physical environment at large-scale in AFGL 5157. 
Section~\ref{sec:disc} deals with the possible SF processes operational in our selected target site. 
Finally, Section~\ref{sec:conc} gives the main conclusions of this paper.
\section{Data}
\label{sec:obser}
Our present work deals with an area of $\sim$0$\degr$.43 (13.5 pc) $\times$ 0$\degr$.43 (13.5 pc)
(central coordinates: {\it l} = 176$\degr$.445; {\it b} = 0$\degr$.215) containing AFGL 5157. 
We analyzed the data sets in the optical, infrared, sub-mm, and radio regimes, which were obtained 
from several surveys. Table~\ref{utab1a} lists all these surveys.

Five College Radio Astronomy 
Observatory (FCRAO) $^{12}$CO(1-0) and $^{13}$CO(1-0) line data (velocity resolution $\sim$0.25~km\,s$^{-1}$) 
were employed in this paper, which were observed as part of the E-OGS. The E-OGS is known as an extended coverage of the FCRAO Outer Galaxy Survey \citep[OGS;][]{heyer98}. 
Typical rms values for the spectra are 0.25 K for $^{12}$CO and 0.2 K for $^{13}$CO \citep[e.g.,][]{heyer96}. 
The CO data cubes were given by M. Heyer and C. Brunt (through private communication). 
The data cubes of $^{12}$CO and $^{13}$CO were smoothed using a Gaussian function with a full-width at half-maximum of 3 pixels to improve sensitivity. 

We also downloaded the {\it Spitzer} post-Basic Calibrated Data (PBCD) images at 3.6, 5.8, and 8.0 $\mu$m from the {\it Spitzer} 
Heritage Archive (ID: 50244; PI: G. Fazio). These images (plate scale $\sim$0$''$.6/pixel) were available only for an area of $\sim$5$'$.5 $\times$ $\sim$5$'$.5 hosting AFGL 5157.
\section{Analysis and Results}
\label{sec:data}
\subsection{Multi-band picture of AFGL 5157}
\label{subsec:u1}
In order to probe the physical environment of AFGL 5157, multi-wavelength continuum images are carefully examined, which span 
from optical H$\alpha$, infrared to radio wavelengths. 
\subsubsection{Multi-wavelength continuum images}
\label{subsec:xu1a}
\begin{figure*}
\epsscale{0.8}
\plotone{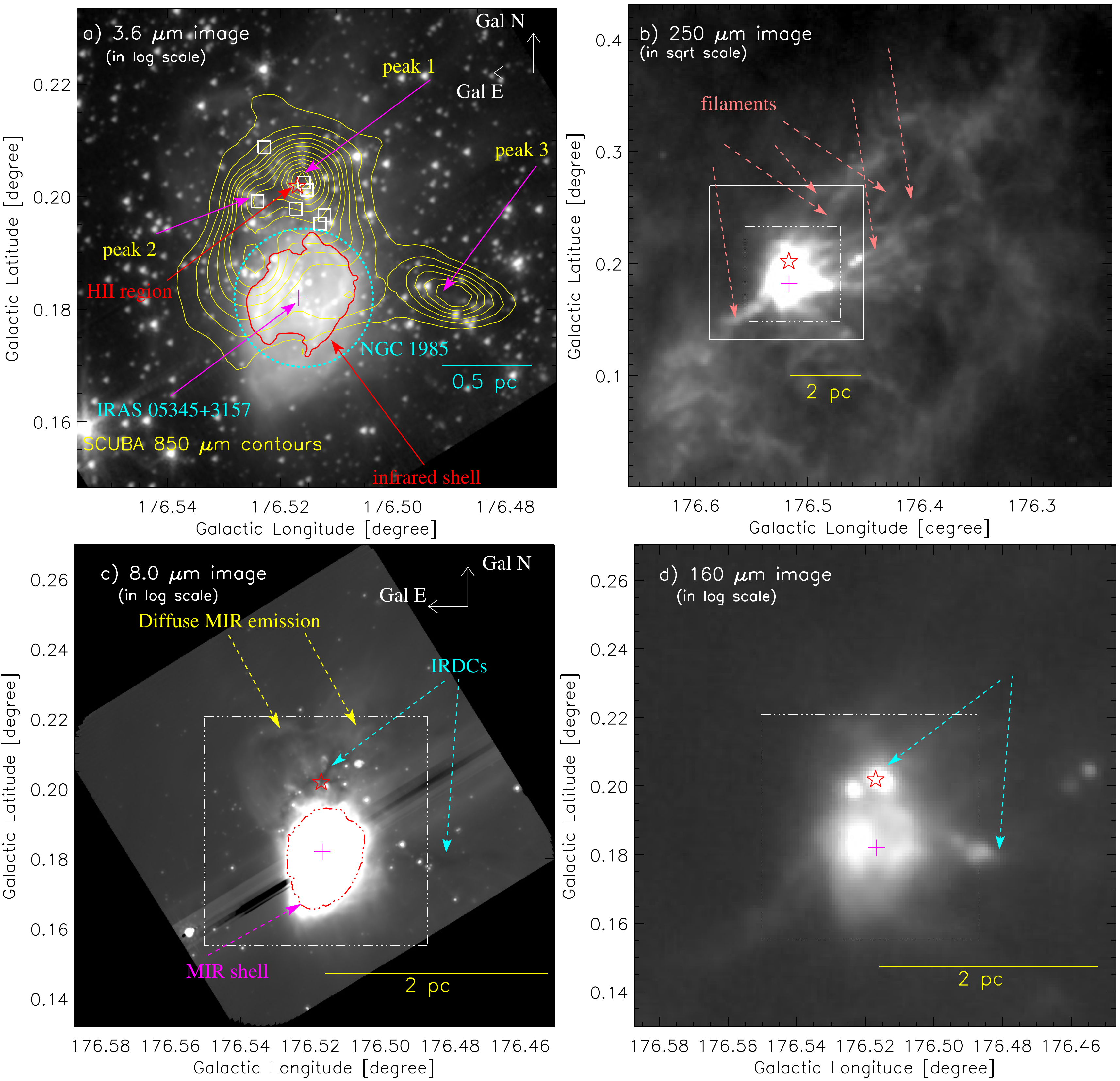}
\epsscale{0.8}
\plotone{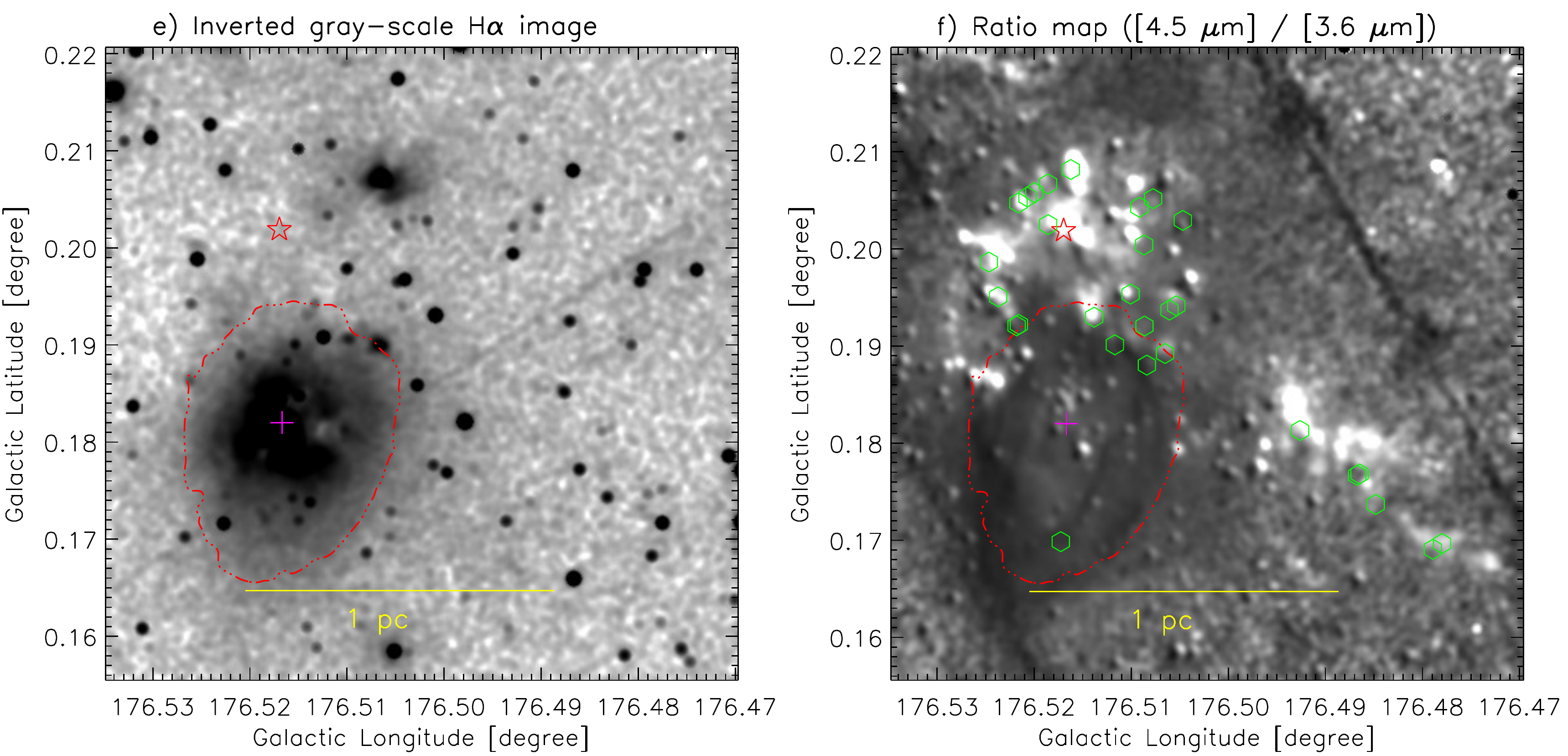}
\caption{Infrared view of AFGL 5157 at different spatial scales. a) {\it Spitzer} 3.6 $\mu$m image (size $\sim$5$'$.1 $\times$ 5$'$.1 or 
$\sim$2.7 pc $\times$ 2.7 pc (at a distance of 1.8 kpc)). 
The SCUBA 850 $\mu$m contours \citep[in yellow;][]{difrancesco08} are shown with the levels 
of 2.64 Jy/beam $\times$ (0.05, 0.1, 0.15, 0.2, 0.25, 0.3, 0.4, 0.5, 0.6, 0.7, 0.8, 0.9, and 0.95). 
The positions of the identified CS cores \citep[from][]{lee11} are also marked in the figure 
(see white squares). Using the {\it Spitzer} 3.6 $\mu$m continuum 
emission contour (in red; level = 7.8 MJy/sr), an infrared shell-like feature is indicated in the figure. 
The area shown in this figure is marked by a broken box in Figure~\ref{fig1}b. 
b) A large-scale view of AFGL 5157 at {\it Herschel} 250 $\mu$m (size $\sim$0$\degr$.43 (13.5 pc) $\times$ 0$\degr$.43 (13.5 pc); 
central coordinates: {\it l} = 176$\degr$.445; {\it b} = 0$\degr$.215). The solid box (in white) encompasses the area shown in Figure~\ref{fig1}c.
c) {\it Spitzer} image at 8.0 $\mu$m (see a solid box in Figure~\ref{fig1}b). 
IRDCs, infrared shell, and diffused emissions are highlighted by arrows. 
The broken box (in white) encompasses the area shown in Figures~\ref{fig1}e and~\ref{fig1}f. 
d) {\it Herschel} image at 160 $\mu$m (see a solid box in Figure~\ref{fig1}b). 
e) Inverted gray-scale IPHAS H$\alpha$ image (see a broken box in Figure~\ref{fig1}d). 
f) {\it Spitzer} ratio map of 4.5 $\mu$m/3.6 $\mu$m emission (see a broken box in Figure~\ref{fig1}d). 
Hexagons (in green) show the locations of MHOs reported by \citet{wolf17} (see also Figure~6b in their paper). 
The H$\alpha$ image and the ratio map are smoothed using a Gaussian function with radius of four (see panels ``e" and ``f"). 
In each panel, a plus symbol shows the location of the IRAS 05343+3157 source. 
In the panels, a scale bar (at a distance of 1.8 kpc) is displayed, and 
a star symbol represents the location of an H\,{\sc ii} region \citep[from][]{torrelles92b}. 
The infrared shell-like feature is highlighted by the {\it Spitzer} 8 $\mu$m contour (in red; level = 70 MJy/sr) in the panels ``c", ``e", and ``f".} 
\label{fig1}
\end{figure*}

As highlighted in Section~\ref{sec:intro}, in the direction of AFGL 5157, Figure~\ref{fig1}a summarizes previously reported features (i.e., infrared shell, dense core, and H\,{\sc ii} region) within an area 
of $\sim$1.8 pc $\times$ $\sim$1.8 pc. A large-scale view of AFGL 5157 (size $\sim$13.5 pc $\times$ 13.5 pc) at 250 $\mu$m is shown in Figure~\ref{fig1}b. This sub-mm image shows the existence of embedded filaments. We highlight representative filaments using arrows in Figure~\ref{fig1}b. The embedded filaments appear to be directed towards the location of AFGL 5157 and the emission peaks (i.e., H\,{\sc ii} region and dense core) in and around the region.  
In Figure~\ref{fig1}b, the position of the previously detected H\,{\sc ii} region \citep[from][]{torrelles92b} is marked by a star, 
which is embedded in the dense NH$_{3}$ core reported by \citet{torrelles92b}. The quantitative information of these filaments is presented in Section~\ref{ssubsec:u2}.

Using the {\it Spitzer} 8.0 $\mu$m image (resolution $\sim$2$''$), Figure~\ref{fig1}c shows a zoomed-in view of the area containing AFGL 5157. 
In Figure~\ref{fig1}c, the location of the infrared shell is also indicated by the emission contour at 8.0 $\mu$m (see also Figure~\ref{fig1}a). 
Furthermore, the absorption features against the Galactic background in the 8.0 $\mu$m image are 
seen in the Galactic northeast to southwest directions around the infrared shell. 
It enables us to identify the infrared dark clouds (IRDCs) in AFGL 5157 (see arrows in Figure~\ref{fig1}c). 
In addition to the IRDCs and the infrared shell, the {\it Spitzer} 8.0 $\mu$m image also reveals noticeable diffuse emission, which is highlighted by arrows in Figure~\ref{fig1}c. 
In Figure~\ref{fig1}d, the sub-mm image at 160 $\mu$m (resolution $\sim$12$''$) traces the bright 
emission toward the IRDCs, and the infrared shell (see arrows in Figure~\ref{fig1}d). 
One can also note that the sub-mm 850 $\mu$m continuum peaks (i.e., peak1, peak2, and peak3) are also found toward the IRDCs (see Figures~\ref{fig1}a and~\ref{fig1}c).

Figure~\ref{fig1}e displays the inverted gray-scale IPHAS H$\alpha$ 
image. 
The locations of the infrared shell, the H\,{\sc ii} region, and IRAS 05345+3157 are also indicated in Figure~\ref{fig1}e. 
The diffuse H$\alpha$ emission is known as a tracer of the ionized emission, which is distributed within the infrared shell powered by massive stars located in the evolved infrared cluster 
\citep[e.g.,][]{chen03}. 
In Figure~\ref{fig1}f, we display the {\it Spitzer} ratio map of 4.5 $\mu$m/3.6 $\mu$m emission, indicating 
the presence of bright and black regions. 
More information on the procedures for obtaining the {\it Spitzer} ratio map can be found in \citet{dewangan17}. 
The positions of previously reported Molecular Hydrogen emission-line Objects \citep[MHOs; from][]{wolf17} are also shown in the figure (see hexagons in Figure~\ref{fig1}f). 
In the ratio map, the bright regions show the excess of 4.5 $\mu$m emission, and are not associated with the 
diffuse H$\alpha$ emission.
On the other hand, the domination of 3.6 $\mu$m emission is inferred though the 
black or dark gray regions seen in the ratio map. 
One can note that polycyclic aromatic hydrocarbon (PAH) emission at 3.3 $\mu$m is included in the IRAC 3.6 $\mu$m band, and a prominent molecular 
hydrogen line emission ($\nu$ = 0--0 $S$(9); 4.693 $\mu$m) is covered in the IRAC 4.5 $\mu$m band. 
Hence, the regions with the excess of 4.5 $\mu$m emission (i.e., bright regions) appear 
to trace the outflow activities in AFGL 5157, which are distributed toward the IRDCs. 
This argument is supported by the fact that the bright regions in the ratio map are spatially matched with 
the positions of MHOs (see Figure~\ref{fig1}f), which are 
driven by YSOs \citep[e.g.,][]{chen03,wolf17}. We also find that the bright regions in the ratio map and the positions of MHOs are found mainly toward the IRDCs or 
the sub-mm 850 $\mu$m continuum peaks, depicting the locations of ongoing SF in AFGL 5157.  
Taking into account the inclusion of 3.3 $\mu$m PAH feature in the 3.6 $\mu$m band, 
the boundaries of the shell-like feature hosting the diffuse H$\alpha$ emission seem to trace 
photodissociation regions (or photon-dominated regions, or PDRs). 
In the Galactic northern direction, the observed diffuse 8.0 $\mu$m emission also hints the presence of an extended PDRs in the selected site.
It is based on the fact that the {\it Spitzer} band at 8.0 $\mu$m contains PAH features at 7.7 and 8.6 $\mu$m.  
\subsubsection{{\it Herschel} filaments in AFGL 5157}
\label{ssubsec:u2}
\begin{figure*}
\epsscale{0.9}
\plotone{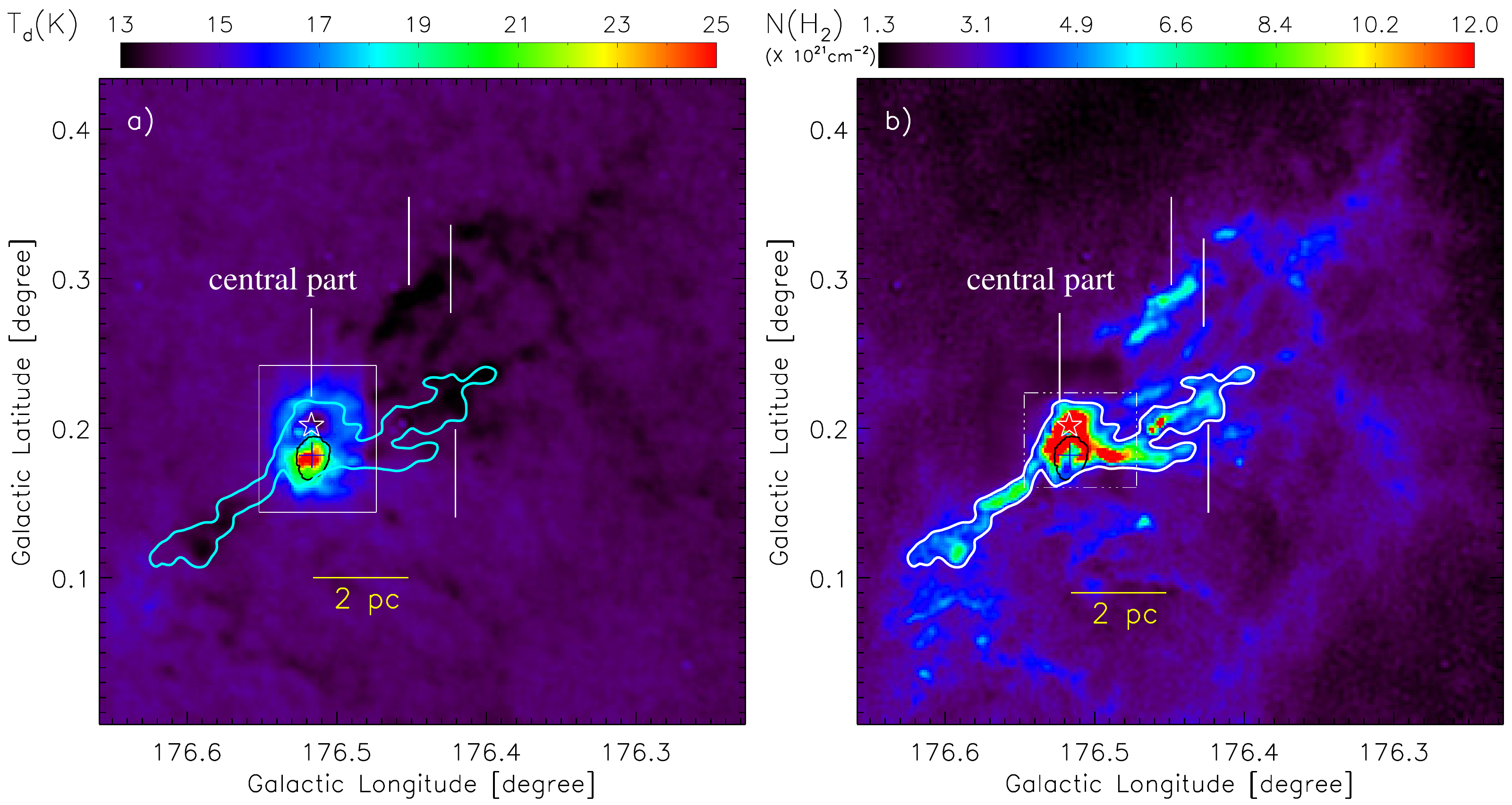}
\epsscale{0.9}
\plotone{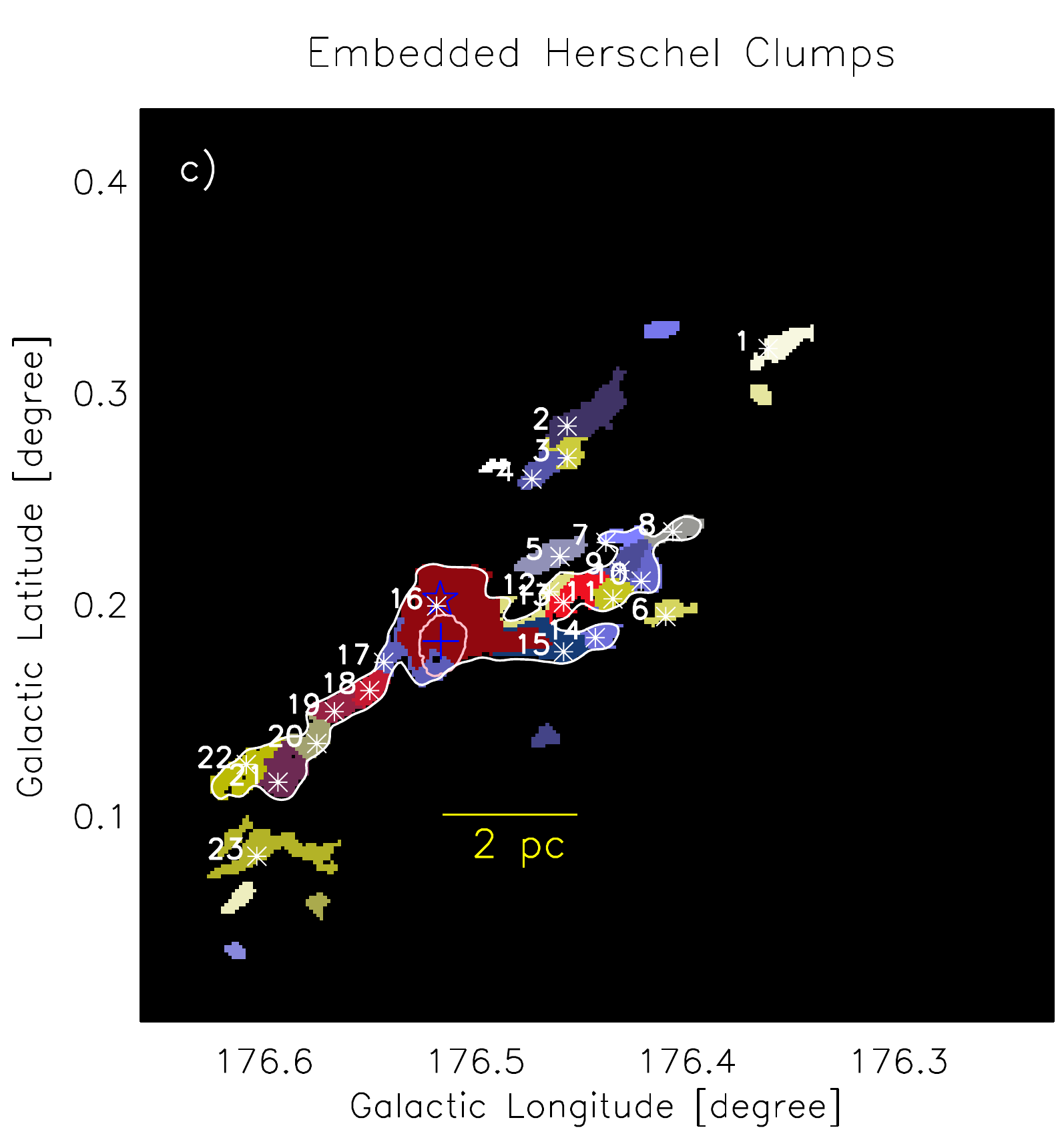}
\caption{a) {\it Herschel} temperature map of the region around AFGL 5157. The solid box (in white) encompasses the area shown 
in Figures~\ref{fig2a}c and~\ref{fig2a}d. 
b) {\it Herschel} column density ($N(\mathrm H_2)$) map. The broken box (in white) encompasses the area 
shown in Figures~\ref{fig2a}a and~\ref{fig2a}b. 
c) The identified clumps are marked by asterisks and the boundary of each {\it Herschel} clump is also 
displayed in the figure. The boundary of each {\it Herschel} clump is shown along with its corresponding clump ID (see also Table~\ref{tab1}). 
In each panel, the filamentary feature is traced in the column density map at a contour level of 3.94 $\times$ 10$^{21}$ cm$^{-2}$. 
In all the panels, other marked symbols and labels are similar to those shown in Figure~\ref{fig1}.}
\label{fig2}
\end{figure*}

In order to further explore the IRDCs, filaments, and infrared shell, we present the {\it Herschel} temperature and column density ($N(\mathrm H_2)$) maps (resolution $\sim$12$''$) in Figures~\ref{fig2}a and~\ref{fig2}b, respectively. 
We retrieved the final processed {\it Herschel} temperature and column density ($N(\mathrm H_2)$) maps (resolution $\sim$12$''$) from the publicly available site\footnote[1]{http://www.astro.cardiff.ac.uk/research/ViaLactea/}.
These maps were produced as a part of the EU-funded ViaLactea project \citep{molinari10b}. 
To produce these {\it Herschel} maps, the Bayesian {\it PPMAP} procedure \citep{marsh15} was employed to the {\it Herschel} continuum data by \citet{marsh17}. 

\begin{figure*}
\epsscale{0.9}
\plotone{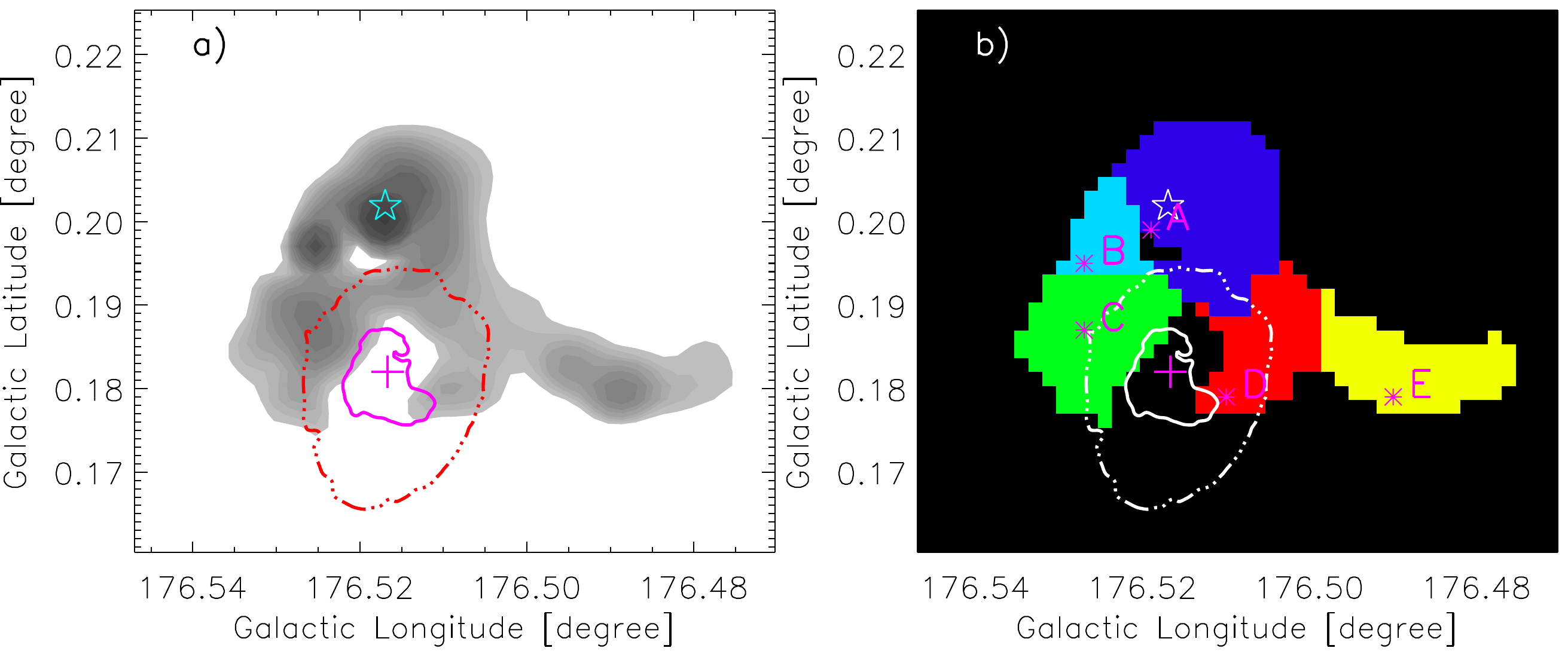}
\epsscale{0.9}
\plotone{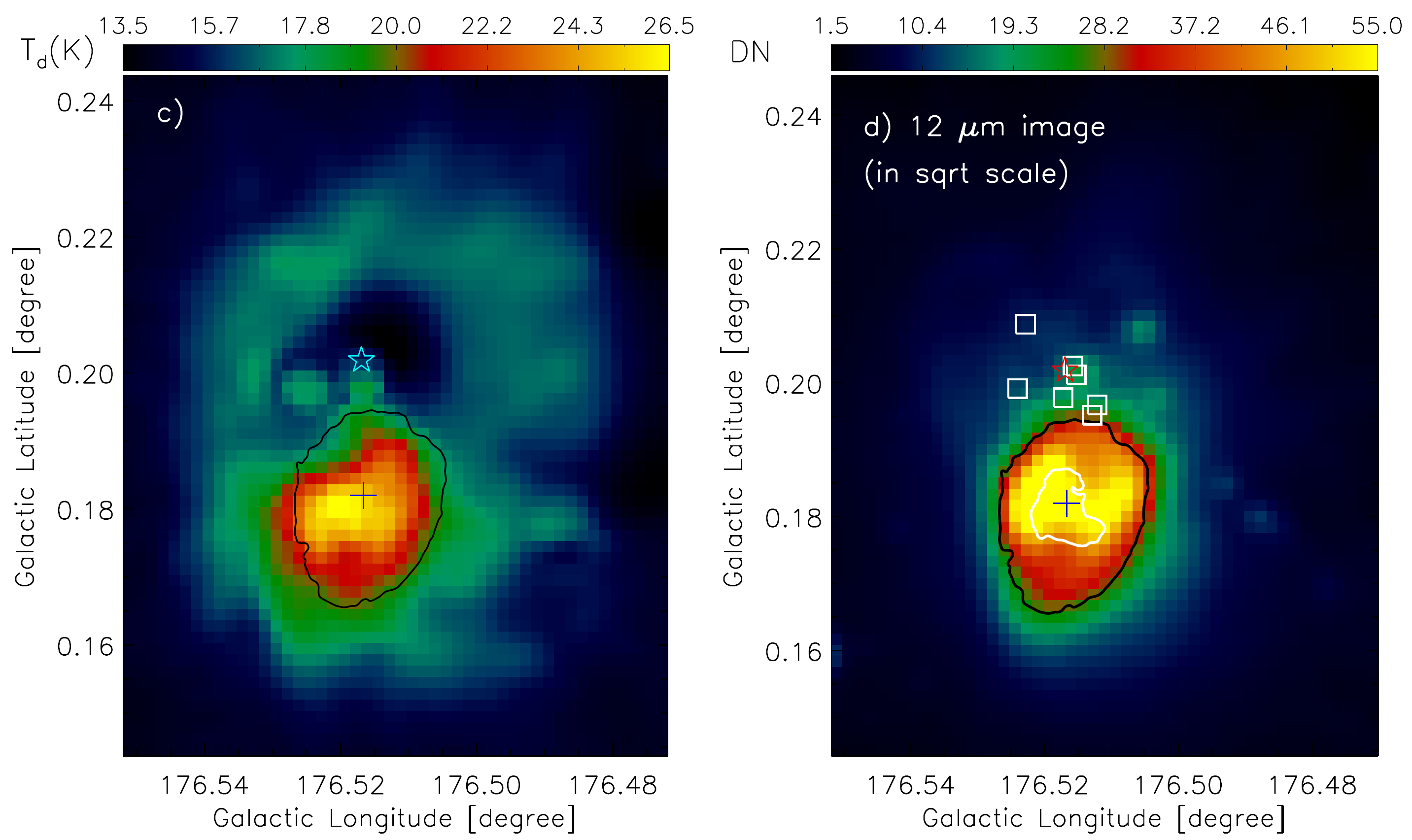}
\caption{a) A zoomed-in view of the central part of the filamentary feature using the {\it Herschel} column density contour map.
The $N(\mathrm H_2)$ contours are shown with the levels of (9, 11, 13, 15, 19, 23, 30, 40, 50, 60, 90, 110, 180, and 300) 
$\times$ 10$^{21}$ cm$^{-2}$. b) Five clumps are marked by asterisks, and are labeled as A, B, C, D, and E in the figure (see also Table~\ref{tab1}). 
c) {\it Herschel} temperature map of an area highlighted by a solid box (in white) in Figure~\ref{fig2}a. 
d) WISE image at 12 $\mu$m of an area highlighted by a solid box (in white) in Figure~\ref{fig2}a. 
The positions of the identified CS cores \citep[from][]{lee11} are also marked in the WISE 12 $\mu$m image (see white squares). 
In the panels ``a" and ``b", a solid contour shows the diffuse H$\alpha$ emission (see also Figures~\ref{fig1}e and~\ref{fig1}f). 
In all the panels, other marked symbols and labels are similar to those shown in Figure~\ref{fig1}.}
\label{fig2a}
\end{figure*}

At least three filaments are highlighted in both the {\it Herschel} maps (see vertical lines in Figures~\ref{fig2}a and~\ref{fig2}b). 
Among these filaments, an elongated filamentary feature (length $\sim$8.3 pc) is identified using a column density contour with a 
level of 3.94 $\times$ 10$^{21}$ cm$^{-2}$, which is marked in Figures~\ref{fig2}a and~\ref{fig2}b. 
The total mass of the filamentary feature is estimated to be $\sim$1170 M$_{\odot}$, and is computed using the equation, $M_{area} = \mu_{H_2} m_H Area_{pix} \Sigma N(H_2)$, where $\mu_{H_2}$ is the mean molecular weight per hydrogen molecule (i.e., 2.8), $Area_{pix}$ is the area subtended by one pixel (i.e., 6$''$/pixel), and 
$\Sigma N(\mathrm H_2)$ is the total column density \citep[see also][]{dewangan17}.  The other two filaments show a temperature range of 
about 13--13.5~K, and seem to be directed toward the central part of the filamentary feature. 
The locations of the infrared shell and the IRDCs are located within the central part of the filamentary feature, which is indicated in the {\it Herschel} maps. 
One can note that a column density deficient region is found in the direction of the infrared shell (see Figure~\ref{fig2}b and also Figure~\ref{fig2a}a).  
The central part of the filamentary feature is traced using a column density contour with the level of $\sim$9--11 $\times$ 10$^{21}$ cm$^{-2}$, where the maximum value of the column density is estimated to be $\sim$3.7 $\times$ 10$^{23}$ cm$^{-2}$. Earlier, \citet{krumholz08} reported a threshold value of 1 gm cm$^{-2}$ (or corresponding 
column density $\sim$3 $\times$ 10$^{23}$ cm$^{-2}$) for formation of massive stars. 
Hence, it is likely that massive stars can be formed within the central part of the filamentary feature. 
Interestingly, \citet{lee11} suggested the presence of massive protostar candidates toward the dense cores. 
The {\it Herschel} temperature map shows an extended temperature structure toward the central part of the filamentary feature. The infrared shell is spatially located inside this 
temperature structure, and is traced in a temperature range of about 20--26.5~K. 
Furthermore, a cold region (T$_{d}$ $\sim$13.5--15~K) inside the temperature structure is also found in the {\it Herschel} temperature map. 
As mentioned earlier, the signatures of outflow activities are mainly found toward the IRDCs, where higher column density materials are observed in the {\it Herschel} column density map. 
Using the column density map and the {\it clumpfind} algorithm \citep{williams94}, we have 
identified twenty three clumps in our 
selected target site, and have also computed thier physical parameters (i.e., mass (M$_{clump}$), radius (R$_{clump}$), and average volume density ($n_{\mathrm H_2}$)). 
To find the clumps, we provided several column density contour levels (i.e., (3.9, 5.0, 6.0, 7.0) $\times$ 10$^{21}$ cm$^{-2}$) as an input parameter for the {\it clumpfind}, where the lowest contour level was chosen at about 6$\sigma$.  
In Figure~\ref{fig2}c, the boundaries of these clumps are shown along with their labels. 
The values of M$_{clump}$, R$_{clump}$, and $n_{\mathrm H_2}$ of each {\it Herschel} clump are listed in Table~\ref{tab1}. 
Here, the average volume density ($n_{\mathrm H_2}$ = 3M$_{clump}$/(4$\pi$R$_{clump}^{3}$$\mu_{H_2} m_H$)) of each clump is estimated using the values of M$_{clump}$ and R$_{clump}$. 
The calculations assume that each clump has a spherical geometry. 
The values of $n_{\mathrm H_2}$ for all the clumps are found between 2200 and 5500 cm$^{-3}$. 
In the direction of the elongated filamentary feature, several clumps are found (see IDs 8--22 in Figure~\ref{fig2}c and also 
Table~\ref{tab1}). Among these clumps, seven clumps (e.g., 13, 15, 16, 17, 19, 21, and 22) have masses 
more or equal to 30 M$_{\odot}$, which are also small scale clumps except the clump ID 16. 
The clump ID 16 is located toward the central part of the filamentary feature (see also Figure~\ref{fig2a}a). 
Additionally, the sub-mm emission toward the clump 13 is prominently seen in the {\it Herschel} images.

Using the {\it Herschel} column density contour map, a zoomed-in view of the central part of the filamentary feature 
is presented in Figure~\ref{fig2a}a. 
As pointed out above, the {\it Herschel} column density map shows no emission toward the region, where the diffuse H$\alpha$ emission (or infrared shell) is observed. 
However, in the direction of clump ID 16, we do not find any column density deficient region due to the choice of a lower column density threshold in the {\it clumpfind}.
Hence, in Figure~\ref{fig2a}a, we have further employed the {\it clumpfind} using higher column density contour levels (i.e., (9, 11, 13, 15, 19, 23, 30, 40, 50, 60) $\times$ 10$^{21}$ cm$^{-2}$). A total of five clumps are identified toward the central part of the filamentary feature (or the clump ID 16), and are labeled as A, B, C, D, and E in the figure. 
The positions of these clumps are marked in Figure~\ref{fig2a}b, and their boundaries are 
also indicated in the figure. In Figure~\ref{fig2a}b, one can also clearly see the column density deficient region toward the diffuse H$\alpha$ 
emission (or infrared shell). Table~\ref{tab1} also lists the values of M$_{clump}$, R$_{clump}$, and $n_{\mathrm H_2}$ of these five clumps. 
The clump masses vary between 45 M$_{\odot}$ and 300 M$_{\odot}$. 
The values of $n_{\mathrm H_2}$ are obtained between $\sim$1.3 $\times$ 10$^{4}$ and $\sim$5.5 $\times$ 10$^{4}$ cm$^{-3}$. 
These dense and massive clumps (i.e., ``A"--``E") are distributed toward the IRDCs.   

Using the {\it Herschel} temperature map, a zoomed-in view of the central part of the filamentary feature is 
also shown in Figure~\ref{fig2a}c. The cold region (T$_{d}$ $\sim$13.5--15~K) is associated with the massive clump ``A", and is 
surrounded by a relatively warm dust emission (T$_{d}$ $\sim$15--18~K). 
In the literature, we find that the clump ``A" has been explored extensively, and is associated with the previously known signatures of ongoing (massive) SF 
(see also the sub-mm 850 $\mu$m continuum peak1 in Figure~\ref{fig1}c). In the direction of the infrared 
shell, the {\it Herschel} column density and temperature maps 
show the low column density tracing the cleared out region associated with the warm dust 
emission (T$_{d}$ $\sim$20--26.5~K) or the evolved infrared cluster. Using the WISE 12 $\mu$m image, Figure~\ref{fig2a}d shows a zoomed-in view of 
the central part, where the positions of the identified seven CS cores \citep[from][]{lee11} are also marked. 
All the CS cores are depicted toward the cold region (T$_{d}$ $\sim$13.5--15~K) traced in the {\it Herschel} temperature map. 
The relatively warmer region (T$_{d}$ $\sim$15--18~K) is seen in the direction of the area, 
where the noticeable diffuse 8.0 and 12.0 $\mu$m emission is evident (see arrows in Figure~\ref{fig1}c).
It seems that the observed extended temperature structure indicates the distribution of the warm dust emission in the PDRs (see Figure~\ref{fig2a}c). 
It implies that the physical environment of AFGL 5157 appears to be affected by the intense energetic feedback of massive stars 
(i.e., stellar wind, ionized emission, and radiation pressure). 
\subsection{Selection and distribution of embedded population in AFGL 5157}
\label{subsec:phot1}
Infrared photometric data have been used to study the embedded YSOs in AFGL 5157.
Using the 2MASS, UKIDSS-GPS, and GLIMPSE360 photometric catalogs, we have utilized 
the dereddened color-color space ([K$-$[3.6]]$_{0}$ and [[3.6]$-$[4.5]]$_{0}$) and the color-magnitude space (H$-$K/K) to 
identify the YSOs. The photometric magnitudes (at 3.6 and 4.5 $\mu$m) were obtained from the Glimpse360 highly reliable catalog. 
We selected only those sources that have photometric magnitude error of 0.2 and less in the 3.6 and 4.5 $\mu$m bands. 
We also utilized the photometric HK data from the UKIDSS GPS tenth archival data release (UKIDSSDR10plus) catalog and the 2MASS. 
Using the conditions given in \citet{lucas08} and \citet{dewangan15}, only reliable UKIDSS GPS photometric data were downloaded in this work. 
2MASS H and K photometric data were also obtained for bright sources that were saturated in the GPS catalog. We considered only those 2MASS sources that have photometric magnitude error of 0.1 
and less in each band.

\begin{figure*}
\epsscale{0.9}
\plotone{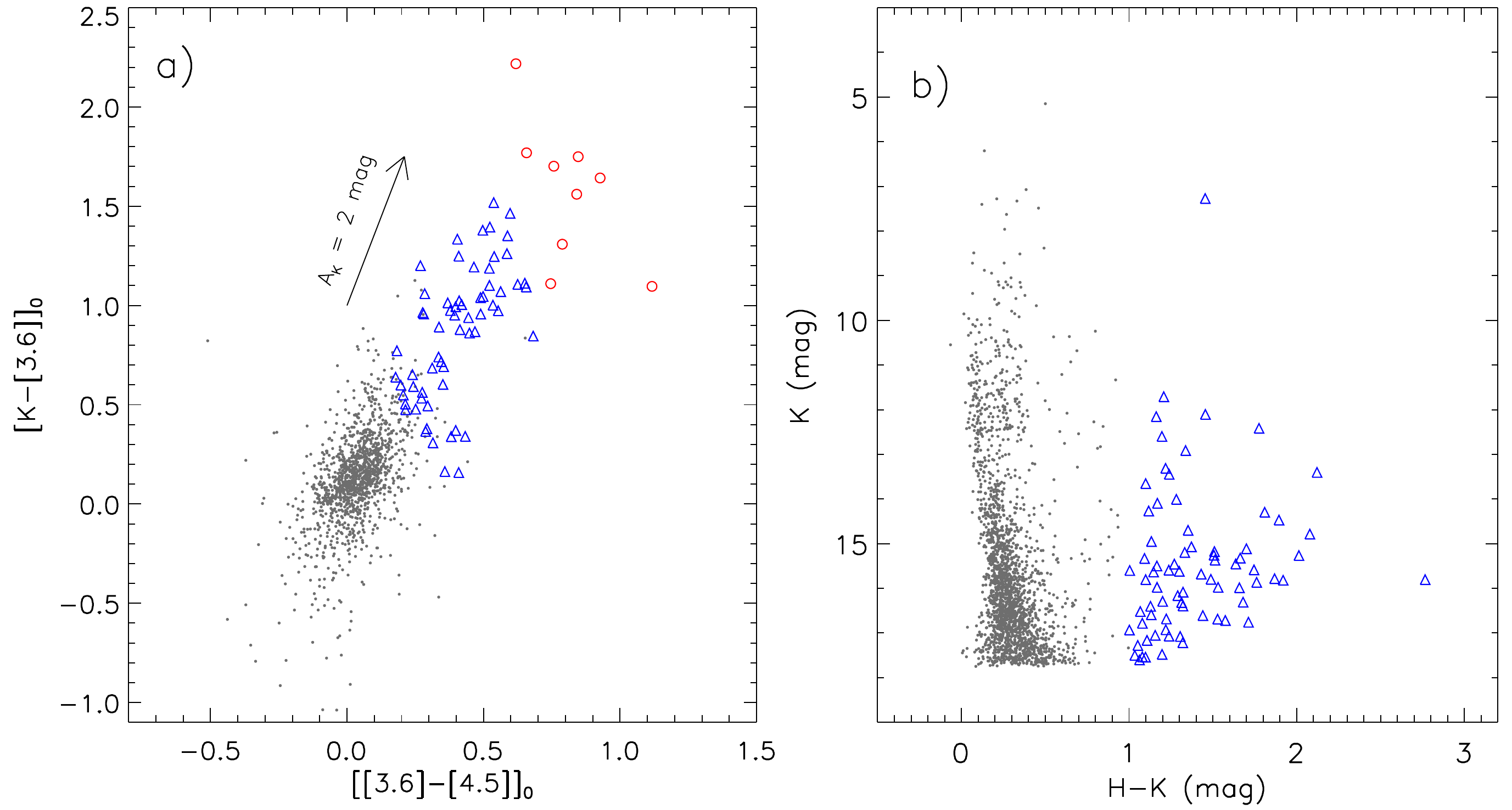}
\epsscale{0.85}
\plotone{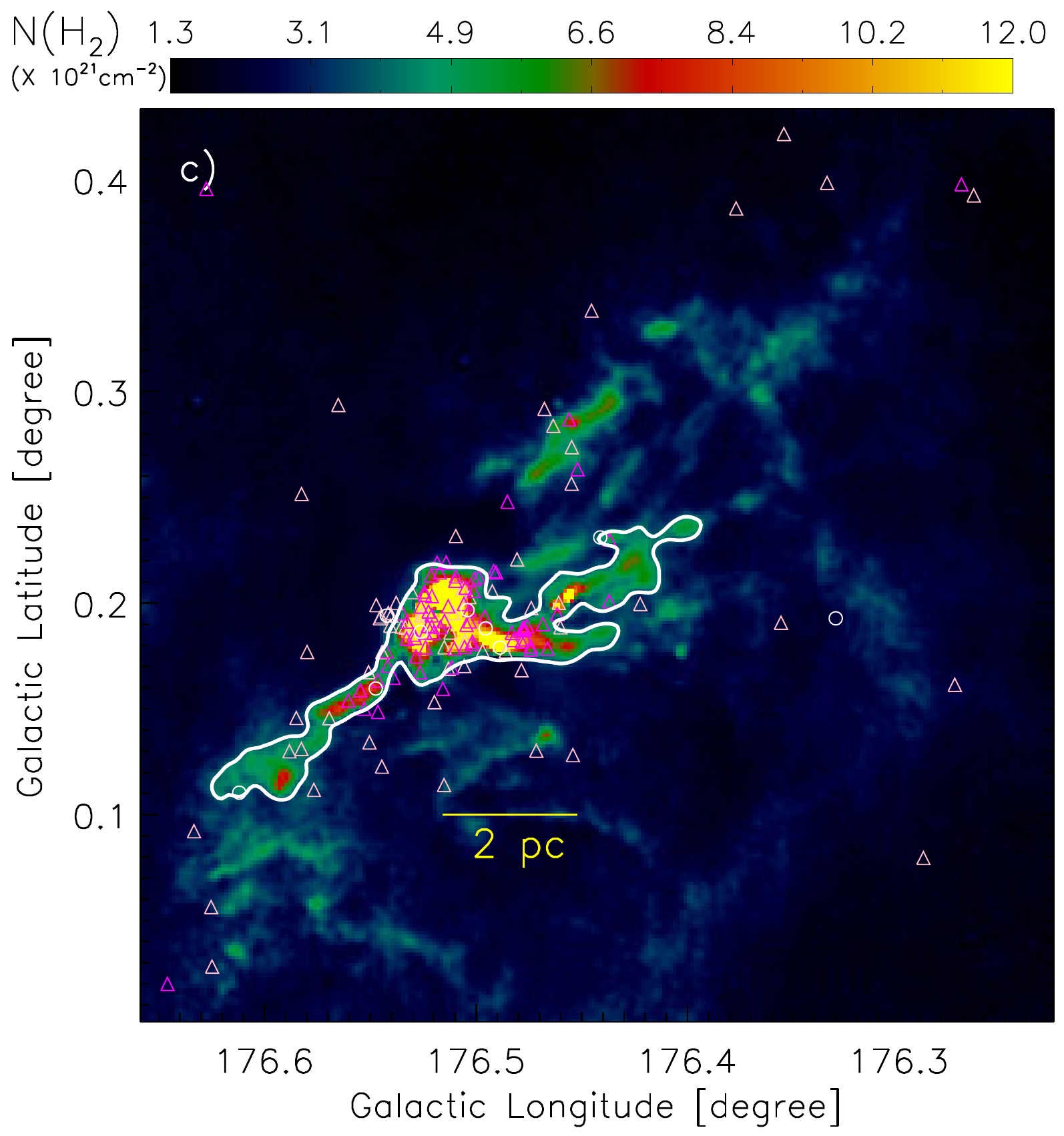}
\caption{a) The panel displays the dereddened [K$-$[3.6]]$_{0}$ $vs$ [[3.6]$-$[4.5]]$_{0}$ color-color plot of sources detected within our selected region (see Figure~\ref{fig1}a). 
The extinction vector \citep[from][]{flaherty07} is shown in the figure. 
b) The panel displays the color-magnitude plot (H$-$K/K) of the sources. 
In the panels ``a" and ``b", Class~I and Class~II YSOs are marked by red circles and open blue triangles, respectively. 
In the panels ``a" and ``b", dot symbols (in gray color) refer stars with only photospheric emission. 
The positions of all the identified YSOs are marked in Figure~\ref{fig9}c.
c) Overlay of the selected YSOs on the {\it Herschel} column density map (see circles and triangles). The filamentary feature is also 
indicated by the column density contour with a level of 3.94 $\times$ 10$^{21}$ cm$^{-2}$ (see also Figure~\ref{fig2}b).}
\label{fig9}
\end{figure*}

\begin{figure*}
\epsscale{1.2}
\plotone{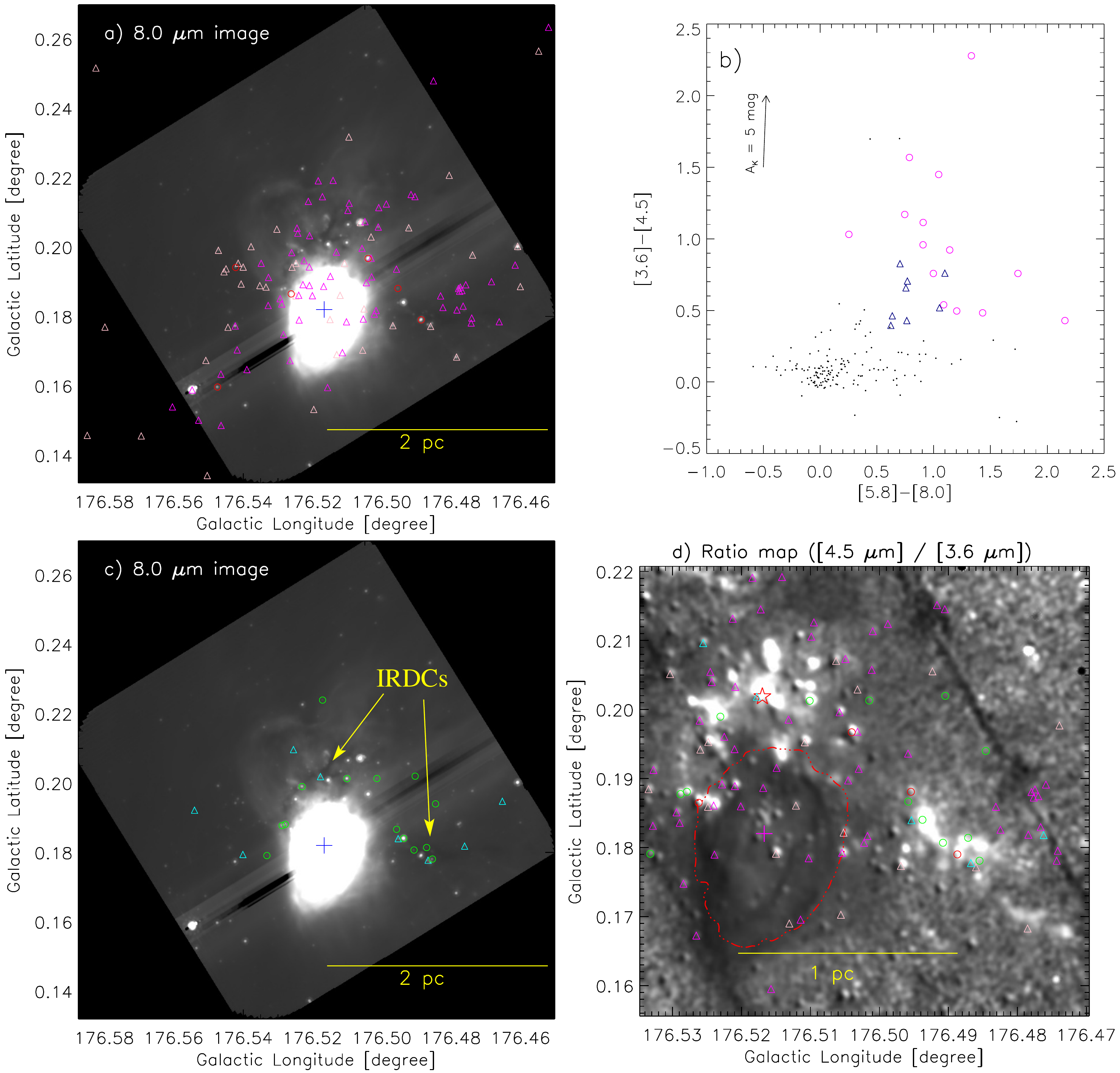}
\caption{a) Overlay of the positions of the selected YSOs on the {\it Spitzer} 8.0 $\mu$m image (see also Figure~\ref{fig9}c). 
b) Color-color plot ([3.6]$-$[4.5] vs. [5.8]$-$[8.0]) of sources detected in the {\it Spitzer} 3.6--8.0 $\mu$m images. 
An extinction vector \citep[from][]{flaherty07} is shown in the figure. Dot symbols (in gray color) show stars with only photospheric emission. 
Circles and triangles show Class~I and Class~II YSOs, respectively. 
c) Overlay of the positions of the additional YSOs on the {\it Spitzer} 8.0 $\mu$m image (from Figure~\ref{fig10}b).
d) Overlay of the positions of all the selected YSOs on the {\it Spitzer} ratio map of 4.5 $\mu$m/3.6 $\mu$m emission (see Figures~\ref{fig10}a and~\ref{fig10}c). The ratio map is the same as in Figure~\ref{fig1}f.}
\label{fig10}
\end{figure*}

Figure~\ref{fig9}a displays the dereddened color-color plot ([K$-$[3.6]]$_{0}$ and [[3.6]$-$[4.5]]$_{0}$) of point-like sources. 
The dereddened colors were computed using the photometric magnitudes of sources at 1--5 $\mu$m and 
the color excess ratios listed in \citet{flaherty07}. 
Following the several conditions suggested in \citet{gutermuth09} \citep[see also][]{dewangan17}, Class~I and Class~II YSOs are identified in our selected site. 
This color-color space gives 71 (9 Class~I and 62 Class~II) YSOs  (see circles and triangles in Figure~\ref{fig9}a). 
Figure~\ref{fig9}b shows the color-magnitude plot (H$-$K/K) of point-like sources.
We find the infrared-excess sources with H$-$K $>$ 1 mag. This color condition is found via the 
color-magnitude analysis of a nearby control field. 
The color-magnitude space gives 74 additional YSO candidates, which are not common with the YSOs identified using 
the dereddened color-color space. These two schemes yield a total of 145 YSOs in the selected site, and their positions are marked in 
the {\it Herschel} column density map (see Figure~\ref{fig9}c). 
A majority of these YSOs are distributed toward the central part of the 
filamentary feature observed in the {\it Herschel} column density map, where five dense clumps A, B, C, D, and E are investigated (see Figures~\ref{fig2a}b and~\ref{fig9}c). 
Furthermore, YSOs are also found toward 
other {\it Herschel} clumps (e.g., 13, 15, 17, 18, 19, and 21; see Figures~\ref{fig2}c and~\ref{fig9}c). 
Overlay of these selected YSOs on the {\it Spitzer} 8.0 $\mu$m image is shown in Figure~\ref{fig10}a.

As previously mentioned that the coverage of {\it Spitzer} 8.0 $\mu$m image (plate scale $\sim$0$''$.6/pixel) is available for an area of $\sim$5$'$.5 $\times$ $\sim$5$'$.5 hosting AFGL 5157. 
Hence, in order to select additional YSOs in this selected area, we have employed the {\it Spitzer} color-color plot ([3.6]-[4.5] vs. [5.8]-[8.0]), 
which is presented in Figure~\ref{fig10}b. This color-color space is useful for identifying the deeply embedded protostars \citep[e.g.,][]{gutermuth09}. 
We extracted photometry of sources at 5.8 and 8.0 $\mu$m, and 
the counterparts of these sources were obtained from the Glimpse360 photometric catalogs at 3.6 and 4.5 $\mu$m. 
Aperture photometry was performed on the {\it Spitzer} images (at 5.8 and 8.0 $\mu$m) with a 2$''$.4 aperture and a sky annulus from 2$''$.4 to 7$''$.3 using IRAF. 
The zero points for these apertures (including aperture corrections) are 17.4899 and 16.6997 mag for the 5.8 and 8.0 $\mu$m bands, 
respectively \citep[see also][for further information on the {\it Spitzer} photometry]{dewangan12}.  
Following the various conditions given in \citet{gutermuth09} \citep[see also][]{dewangan12}, additional YSOs are identified. 
We have classified these selected YSOs into different evolutionary stages based on their 
slopes of the spectral energy distribution ($\alpha_{3.6-8.0}$) computed from 3.6 to 8.0 $\mu$m 
(i.e., Class~I ($\alpha_{3.6-8.0} > -0.3$), Class~II ($-0.3 > \alpha_{3.6-8.0} > -1.6$), 
and Class~III ($-1.6> \alpha_{3.6-8.0} > -2.56$)) \citep[e.g.,][]{lada06}. 
We select additional 22 YSOs (14 Class~I and 8 Class~II) using this scheme, which are 
plotted in Figure~\ref{fig10}c. These selected YSOs are not overlapped with that YSOs shown in Figure~\ref{fig10}a. 
Photometric information of all the selected YSOs is given in Table~\ref{tab3}. 
Figure~\ref{fig10}d shows the overlay of the selected YSOs on the {\it Spitzer} ratio map of 4.5 $\mu$m/3.6 $\mu$m emission, 
revealing the association of YSOs with the bright regions due to an excess of 4.5 $\mu$m emission 
(or outflow activities; see Section~\ref{subsec:u1}). 

\begin{figure*}
\epsscale{1.2}
\plotone{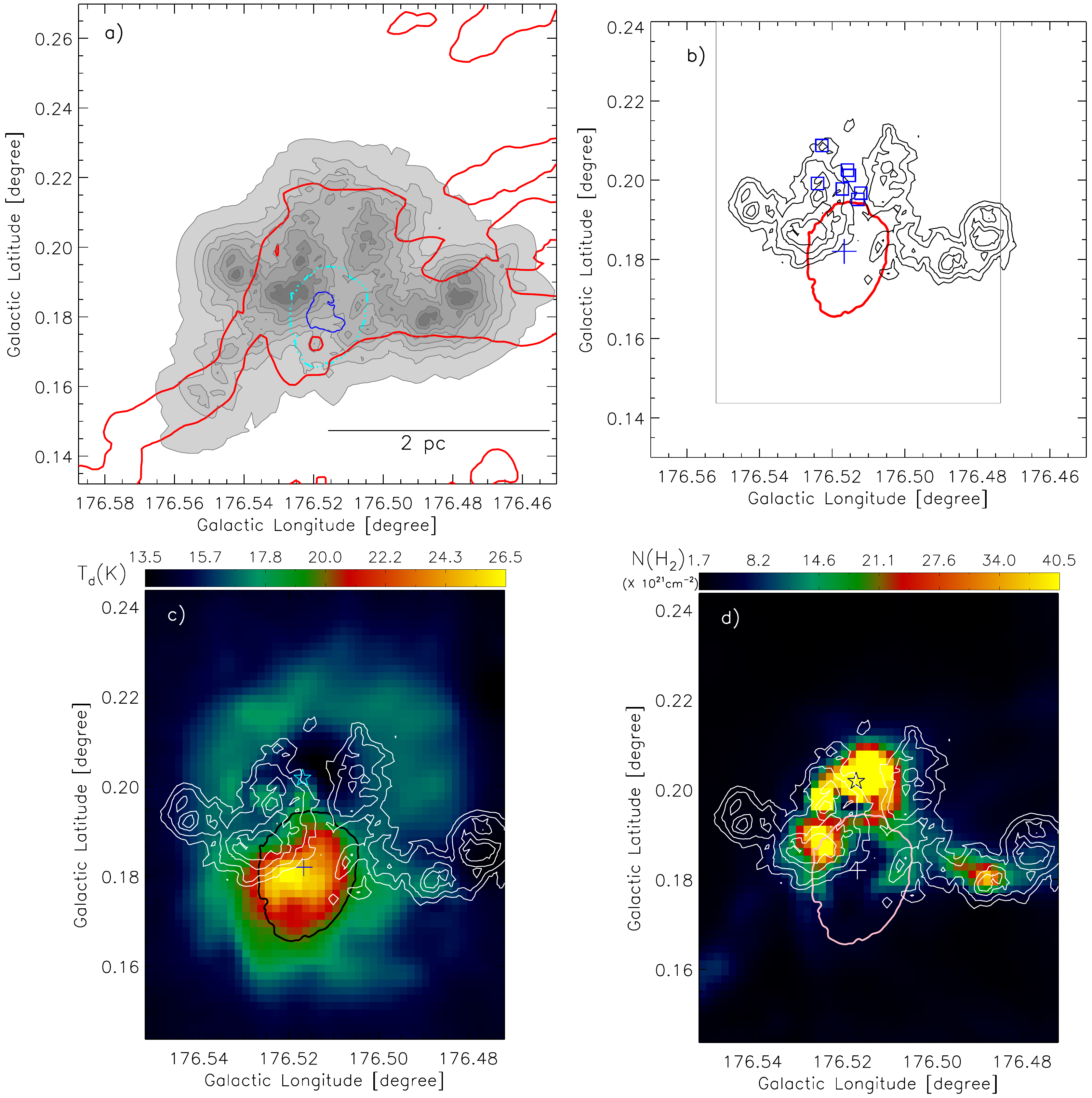}
\caption{a) The panel shows the surface density contour map (in gray scale) of all the identified YSOs (see Figures~\ref{fig10}a and~\ref{fig10}c). The surface density contours are also shown with the levels of 5, 10, 15, 20, 30, 40, 60, and 85 YSOs/pc$^{2}$, from the outer to the inner side. The {\it Herschel} column density contours (in red), the {\it Spitzer} 8.0 $\mu$m emission contour (in cyan), and the H$\alpha$ emission contour (in blue) are also marked in the figure. 
b) Surface density contours of YSOs (in black) against the location of the infrared shell (see red contour). 
Blue squares represent the positions of the identified CS cores \citep[from][]{lee11}. 
c) Overlay of the surface density contours (in white) on the {\it Herschel} temperature map. 
d) Overlay of the surface density contours (in white) on the {\it Herschel} column density map. 
In the last three panels, the surface density contours are shown with the levels of 30, 40, 60, and 85 YSOs/pc$^{2}$. 
In all the panels, the infrared shell is indicated by the {\it Spitzer} 8.0 $\mu$m emission contour as in Figure~\ref{fig1}c.}
\label{fig11}
\end{figure*}

Figure~\ref{fig11}a displays a surface density contour map of 135 YSOs distributed in the area shown in Figure~\ref{fig10}c.
The surface density contours of YSOs are shown with the levels of 5, 10, 15, 20, 30, 40, 60, and 85 YSOs/pc$^{2}$, suggesting 
the intense ongoing SF activities in the selected site. 
In Figure~\ref{fig11}a, the {\it Herschel} column density contours (in red), the infrared shell (in cyan), and the diffuse H$\alpha$ emission contour (in blue) are also highlighted in the surface density contour map. We employed the nearest-neighbour (NN) method to estimate the surface density contours of the selected YSOs \citep[see][for more details]{casertano85,gutermuth09,bressert10,dewangan17}. 
Adopting the similar analysis as carried out in \citet{dewangan17}, the surface density map of all the selected YSOs has been produced 
using a 5$\arcsec$ grid and 6 NN at a distance of 1.8 kpc. 
The surface density contours with the levels of 30, 40, 60, and 85 YSOs/pc$^{2}$ are shown in 
Figures~\ref{fig11}b,~\ref{fig11}c, and~\ref{fig11}d. 
In Figure~\ref{fig11}b, we have also marked the positions of the identified seven CS cores \citep[from][]{lee11}. 
Using the continuum map at 2.7 mm and the CS line data, \citet{lee11} also reported that massive protostar candidates are forming in two of these dense cores.   
In Figures~\ref{fig11}c and~\ref{fig11}d, the surface density contours are overlaid on the {\it Herschel} column density and temperature maps, respectively. 
Figure~\ref{fig11}d reveals a spatial correlation between the embedded clusters (with surface density $\geq$ 30 YSOs/pc$^{2}$) 
and the higher column density materials (including  {\it Herschel} clumps and dense CS cores). 

Together, the central part of the filamentary feature is associated with massive dust clumps, dense CS cores, massive protostar candidates, and clusters of YSOs. 
\subsection{Kinematics of molecular gas}
\label{sec:coem} 
\begin{figure*}
\epsscale{0.44}
\plotone{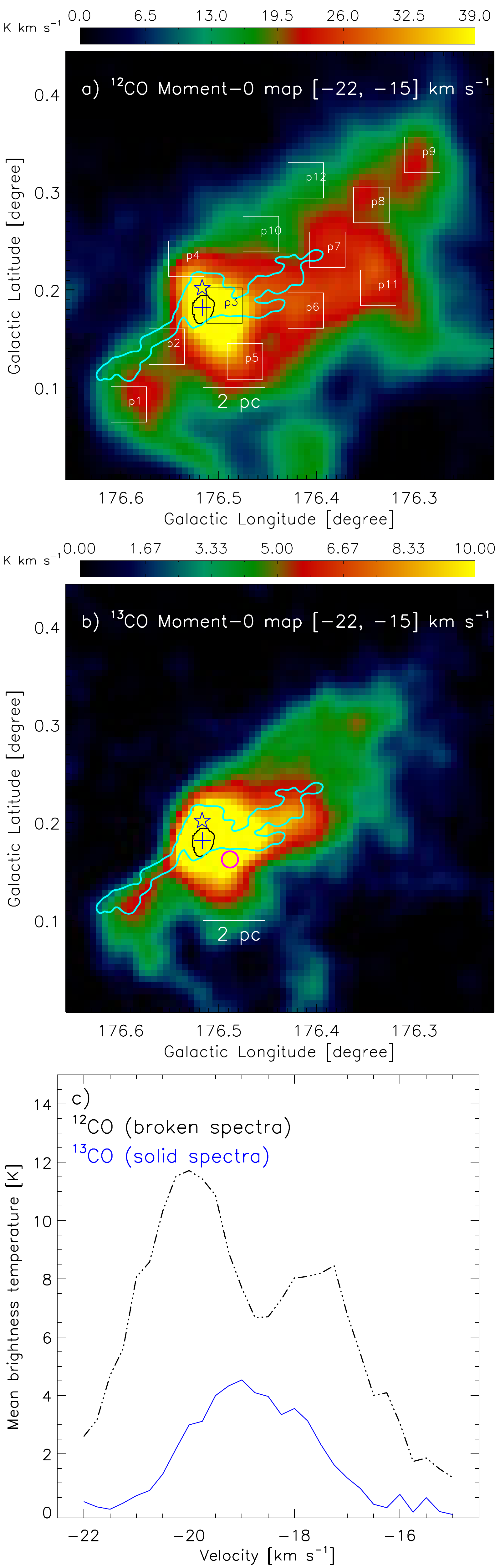}
\caption{a) $^{12}$CO(J =1$-$0) map of intensity (moment-0), integrated from $-22$ to $-$15 km s$^{-1}$. 
Twelve small regions (i.e. p1 to p12) are also indicated by boxes in the figure.  
b) $^{13}$CO(J =1$-$0) map of intensity (moment-0), integrated from $-22$ to $-$15 km s$^{-1}$.
In the panels ``a" and ``b", the {\it Herschel} column density contour (in cyan) is also overlaid on 
the molecular map with a level of 3.94 $\times$ 10$^{21}$ cm$^{-2}$, indicating the location of the {\it Herschel} filamentary feature. 
c) The panel displays the $^{12}$CO and $^{13}$CO spectra. The profiles are obtained by averaging the area shown in Figure~\ref{fig4}b (see a circle in Figure~\ref{fig4}b). 
In the panels ``a" and ``b", other marked symbols and labels are similar to those shown in Figure~\ref{fig1}.}
\label{fig4}
\end{figure*}

\begin{figure*}
\epsscale{1}
\plotone{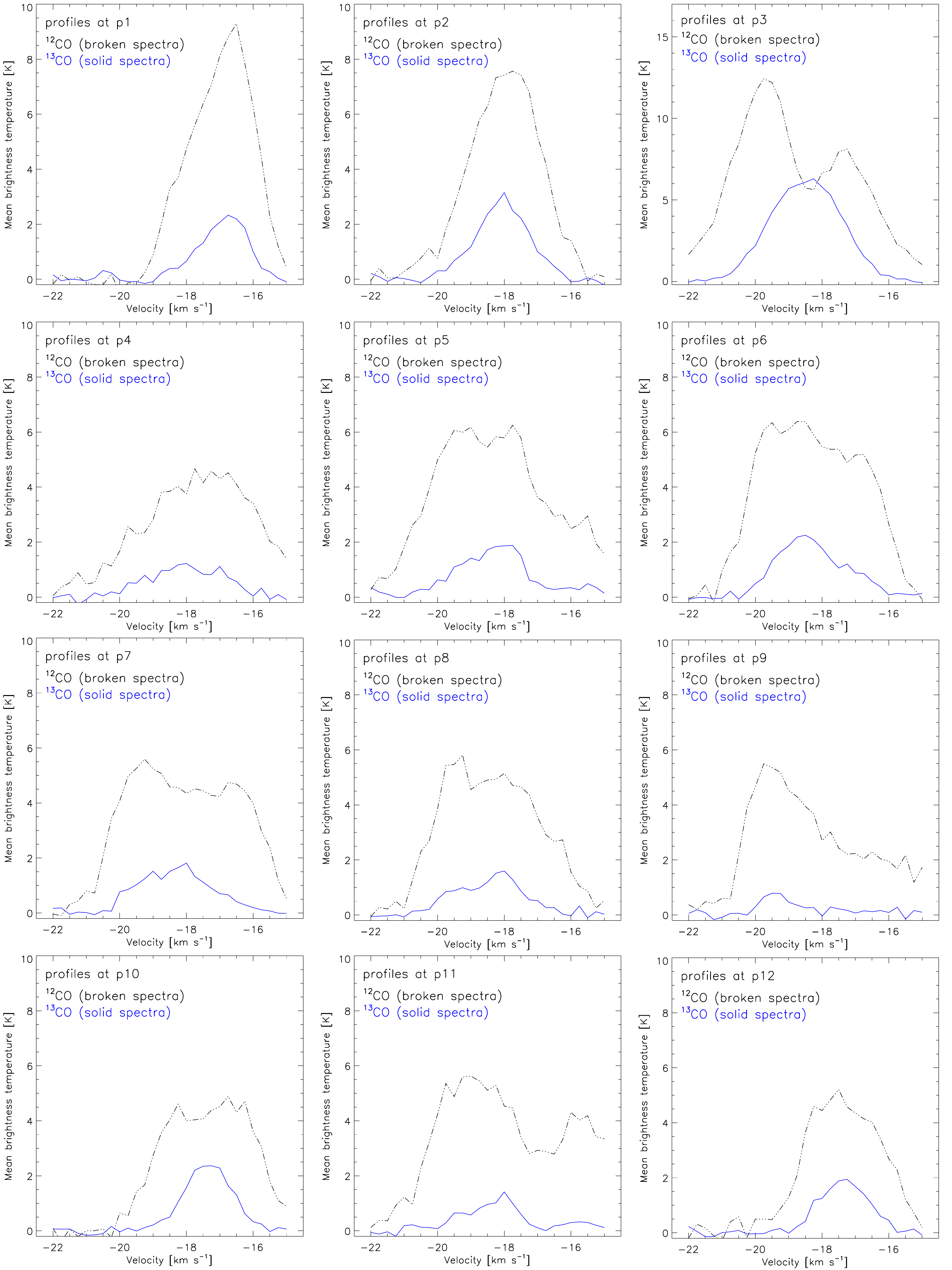}
\caption{The $^{12}$CO and $^{13}$CO profiles in the direction of twelve small regions (i.e. p1 to p12; see corresponding boxes 
in Figure~\ref{fig4}a).}
\label{ufig4}
\end{figure*}

To study the molecular gas associated with AFGL 5157, Figures~\ref{fig4}a and~\ref{fig4}b show 
the $^{12}$CO(J =1$-$0) and $^{13}$CO(J =1$-$0) intensity maps (moment-0) integrated over a velocity 
range of $-$22 to $-$15 km s$^{-1}$, respectively. The locations of the filamentary feature, infrared shell, and previously known 
H\,{\sc ii} region are highlighted in both the molecular maps. We find that all the {\it Herschel} filaments are embedded in the molecular cloud associated with AFGL 5157. 
Note that our molecular line data, having coarse resolutions, do not 
allow to trace the gas motions along/into the filamentary structures seen in the {\it Herschel} data. 
However, these molecular line data enable us to study the large-scale molecular cloud motions. 
The observed $^{12}$CO and $^{13}$CO profiles are presented in Figure~\ref{fig4}c. 
The spectra are produced by averaging the area highlighted by a solid circle in Figure~\ref{fig4}b, 
which is located toward the the central part of the filamentary feature. 
The $^{12}$CO profile reveals two velocity peaks (around $-$20 and $-$17 km s$^{-1}$), while 
a broad profile of $^{13}$CO is observed toward our selected area (see a solid circle in Figure~\ref{fig4}b). 
Figure~\ref{ufig4} displays the average $^{12}$CO and $^{13}$CO spectra in the direction of twelve small regions (i.e., p1 to p12; 
see corresponding boxes in Figure~\ref{fig4}a). 
The selected region ``p3" covering larger area and the small circle in Figure~\ref{fig4}b are very close, where one may expect similar molecular profiles.  
In the $^{12}$CO spectrum, two peaks (around $-$20 and $-$17 km s$^{-1}$) are clearly seen in the direction of the region p3, while 
the peak of the corresponding $^{13}$CO spectrum (around $-$18.5 km s$^{-1}$) is found toward a dip between two peaks seen in the $^{12}$CO spectrum (see also Figure~\ref{fig4}c). In the direction of other small regions (or away from the central part of the filamentary feature), the peak in the $^{12}$CO spectrum nearly matches with the peak in the $^{13}$CO profile. Hence, in the direction of the region p3, the shift of velocity in the $^{12}$CO and $^{13}$CO could be due to the presence of self-absorption of $^{12}$CO line around $-$18.5 km s$^{-1}$. 
It can be treated as an example of a blue-asymmetric self-absorption profile. Based on a relatively low ratio value ($<$ 2.5) of $^{12}$CO/$^{13}$CO at $-$18.5 km s$^{-1}$, 
the $^{12}$CO self-absorption feature is found toward the elongated filamentary feature (see a summary figure in this Section and also Section~\ref{sec:disc} for more details). If we use the ratio value less than one (i.e., $<$ 1.0) then one may notice significant self-absorption in the direction of the {\it Herschel} clump ``D", which is an area between the {\it Herschel} clump ``E" (or the sub-mm peak3) and the infrared shell.

\begin{figure*}
\epsscale{1.15}
\plotone{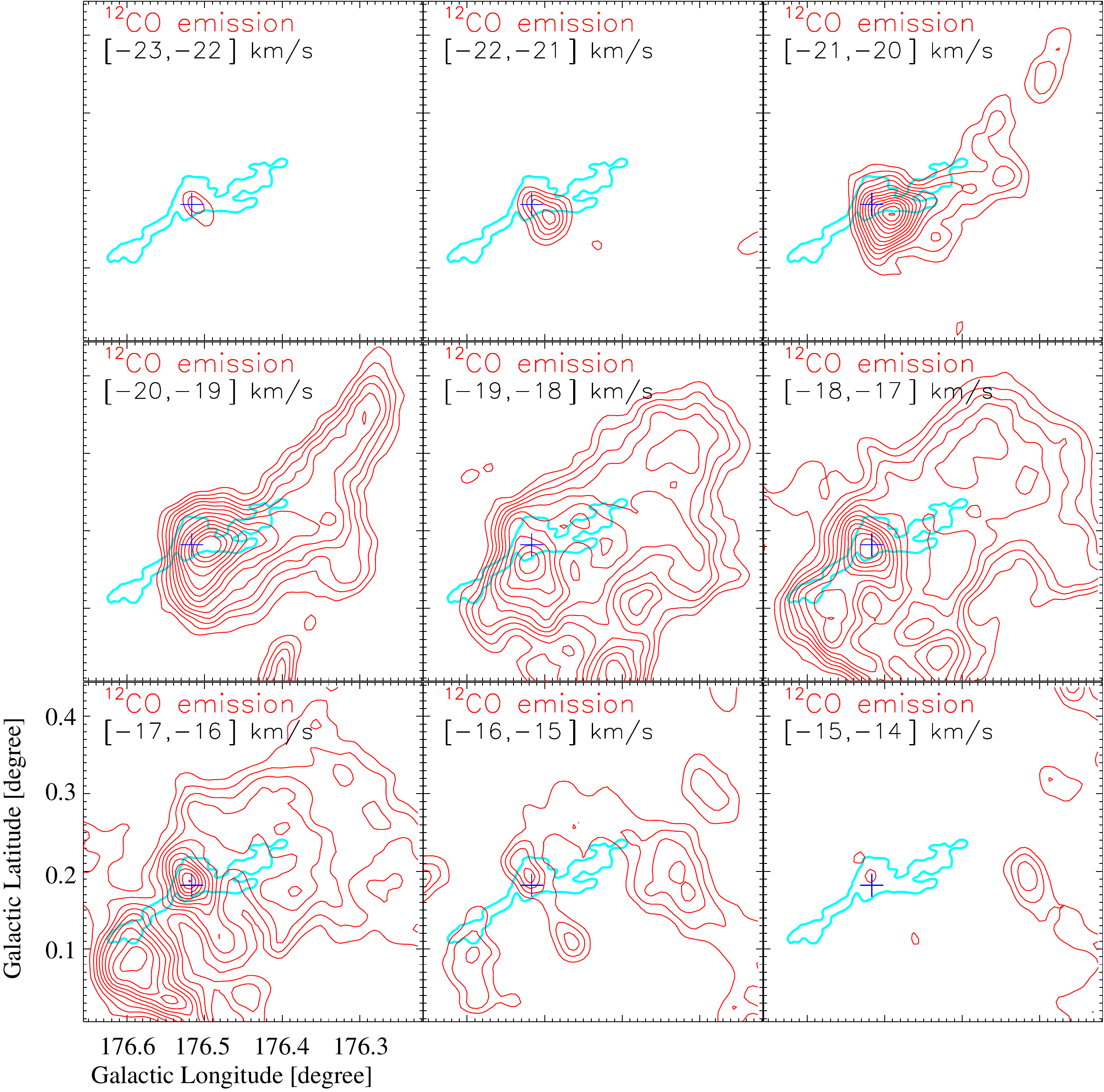}
\caption{Velocity channel contours of $^{12}$CO emission. 
The molecular emission is integrated over a velocity interval, which is marked in each panel (in km s$^{-1}$). 
The contour levels of $^{12}$CO are presented with the levels of 2, 3, 4, 5, 6, 7, 8, 9, 10, 11, 12, 13, 
and 14 K km s$^{-1}$. In each panel, the filamentary feature is indicated by the column density contour with a level of 3.94 $\times$ 10$^{21}$ cm$^{-2}$ (see also Figure~\ref{fig2}b). In all the panels, a plus symbol 
shows the location of the IRAS 05343+3157 source.}
\label{fig5}
\end{figure*}

\begin{figure*}
\epsscale{1.15}
\plotone{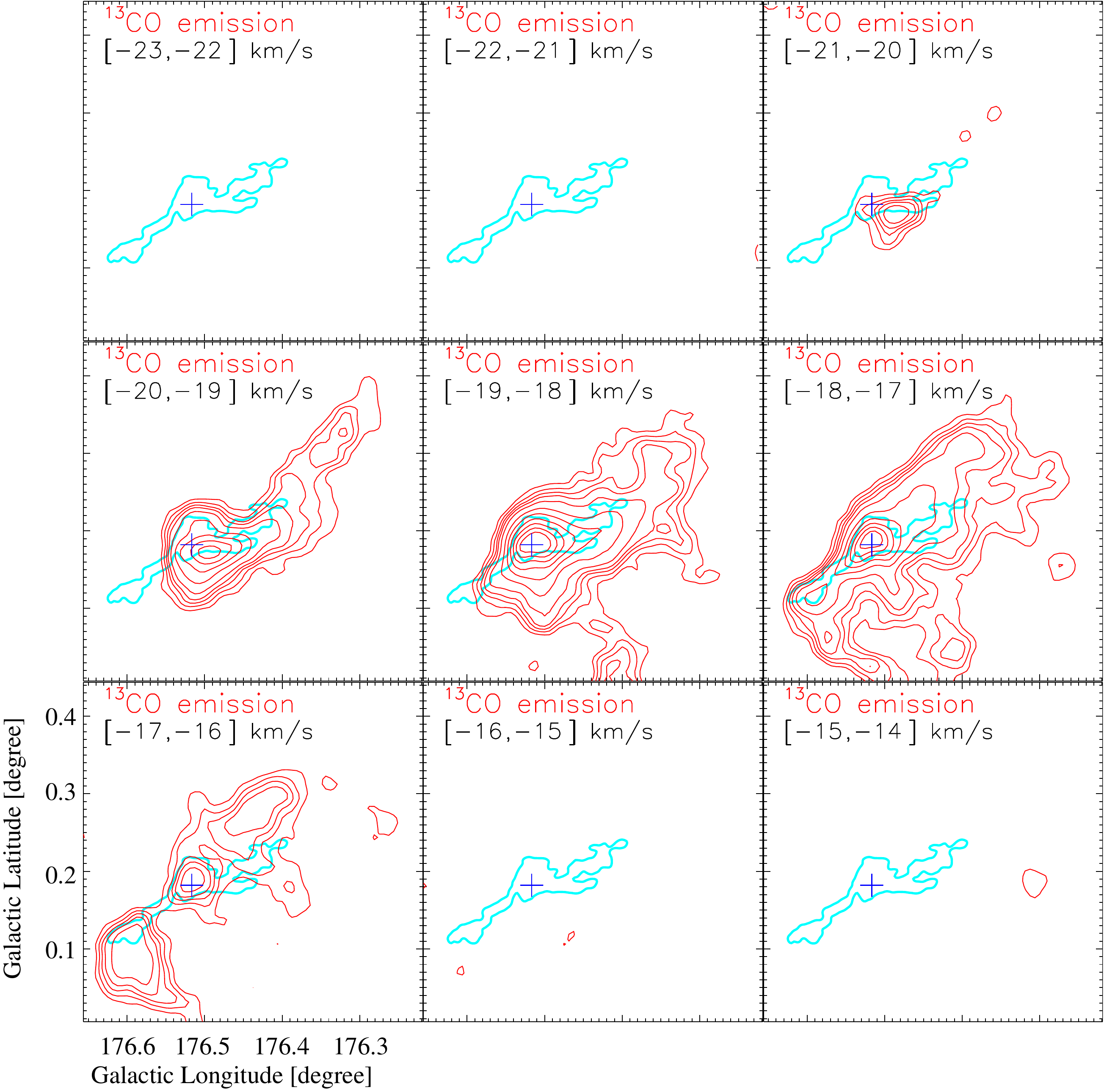}
\caption{Velocity channel contours of $^{13}$CO emission. The contour levels of $^{13}$CO are presented with 
the levels of 0.6, 0.9, 1.2, 1.5, 2, 3, 4, 5, 6, 7, and 8 K km s$^{-1}$. In each panel, other marked symbols and the contour are similar to those shown in Figure~\ref{fig5}.}
\label{fig6}
\end{figure*}

\begin{figure*}
\epsscale{1.15}
\plotone{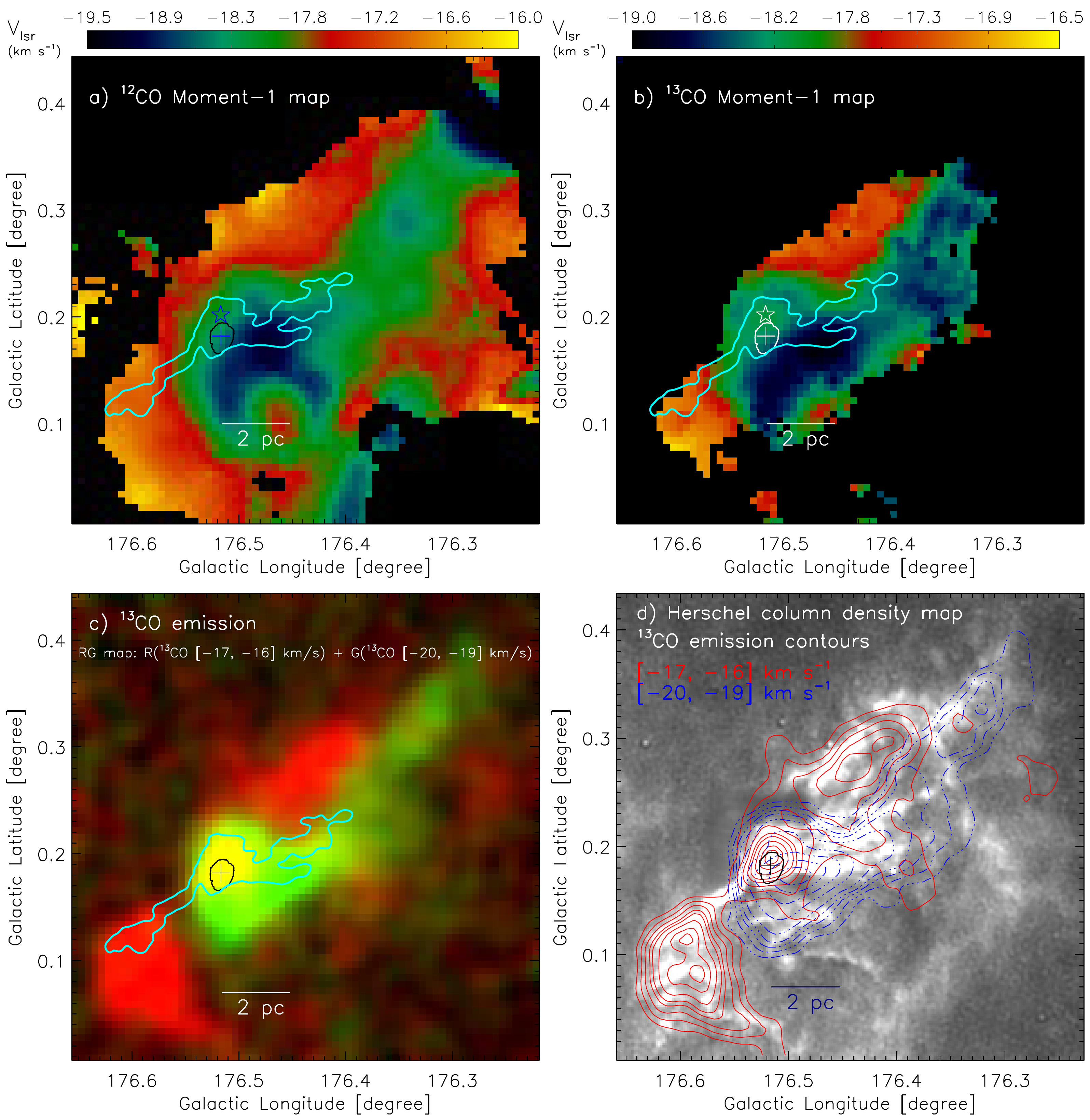}
\caption{a) $^{12}$CO first moment map. 
b) $^{13}$CO first moment map.
c) Two color-composite image produced using the $^{13}$CO maps at [$-$17, $-$16] 
and [$-$20, $-$19] km s$^{-1}$ in red and green, respectively.
d) Overlay of the $^{13}$CO emission contours (at [$-$17, $-$16] and [$-$20, $-$19] km s$^{-1}$) on 
the {\it Herschel} column density map. The velocity ranges are also given in the figure. 
The $^{13}$CO emission contours at [$-$17, $-$16] km s$^{-1}$ (in red) are displayed with the levels of 0.6, 0.9, 1.2, 1.5, 1.7, 2, 2.3, 
and 2.5 K km s$^{-1}$, while the $^{13}$CO emission contours at [$-$20, $-$19] 
km s$^{-1}$ (in blue) are 0.6, 0.9, 1.2, 1.5, 2, 3, 4, 5, 6, 7, and 8 K km s$^{-1}$. 
In each panel, other marked symbols and the contour are similar to those shown in Figure~\ref{fig2}.}
\label{fig7}
\end{figure*}

\begin{figure*} 
\epsscale{0.56}
\plotone{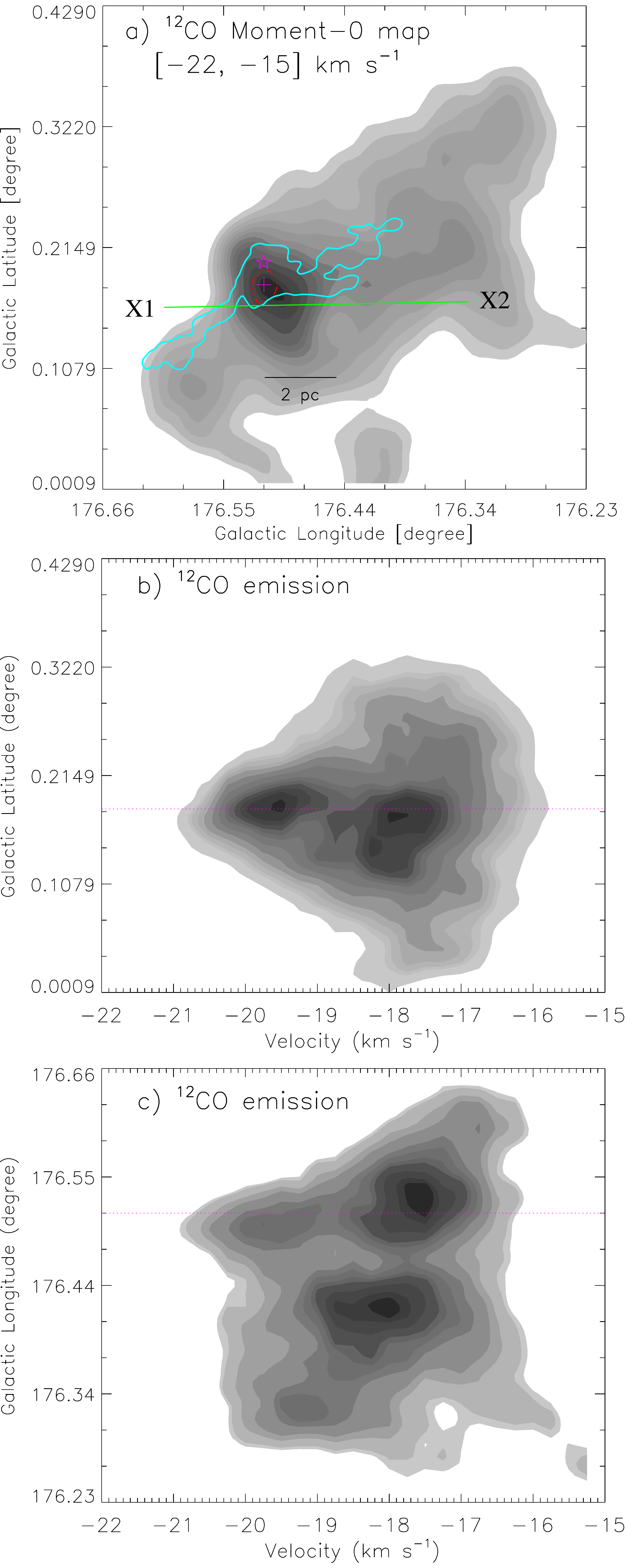}
\caption{a) Contour map of the integrated $^{12}$CO emission in a velocity range of $-$22 to $-$15 km s$^{-1}$. 
The $^{12}$CO integrated intensity map is similar as in Figure~\ref{fig4}a. 
Other marked symbols and contour are similar to those shown in Figure~\ref{fig2}. 
A solid line (in green) shows the axis (i.e., X1--X2), where position-velocity maps are 
obtained (see Figures~\ref{fig8v}a and~\ref{fig8v}b). 
b) Latitude-velocity map of $^{12}$CO. The molecular emission is integrated over the longitude from 176$\degr$.23 to 176$\degr$.66. 
c) Longitude-velocity map of $^{12}$CO. The molecular emission is integrated over the latitude from 0$\degr$.0009 to 0$\degr$.4290. 
In the panels ``b" and ``c", a dotted magenta line shows the position of the IRAS 05343+3157 source.}
\label{fig8}
\end{figure*}

\begin{figure*} 
\epsscale{1.1}
\plotone{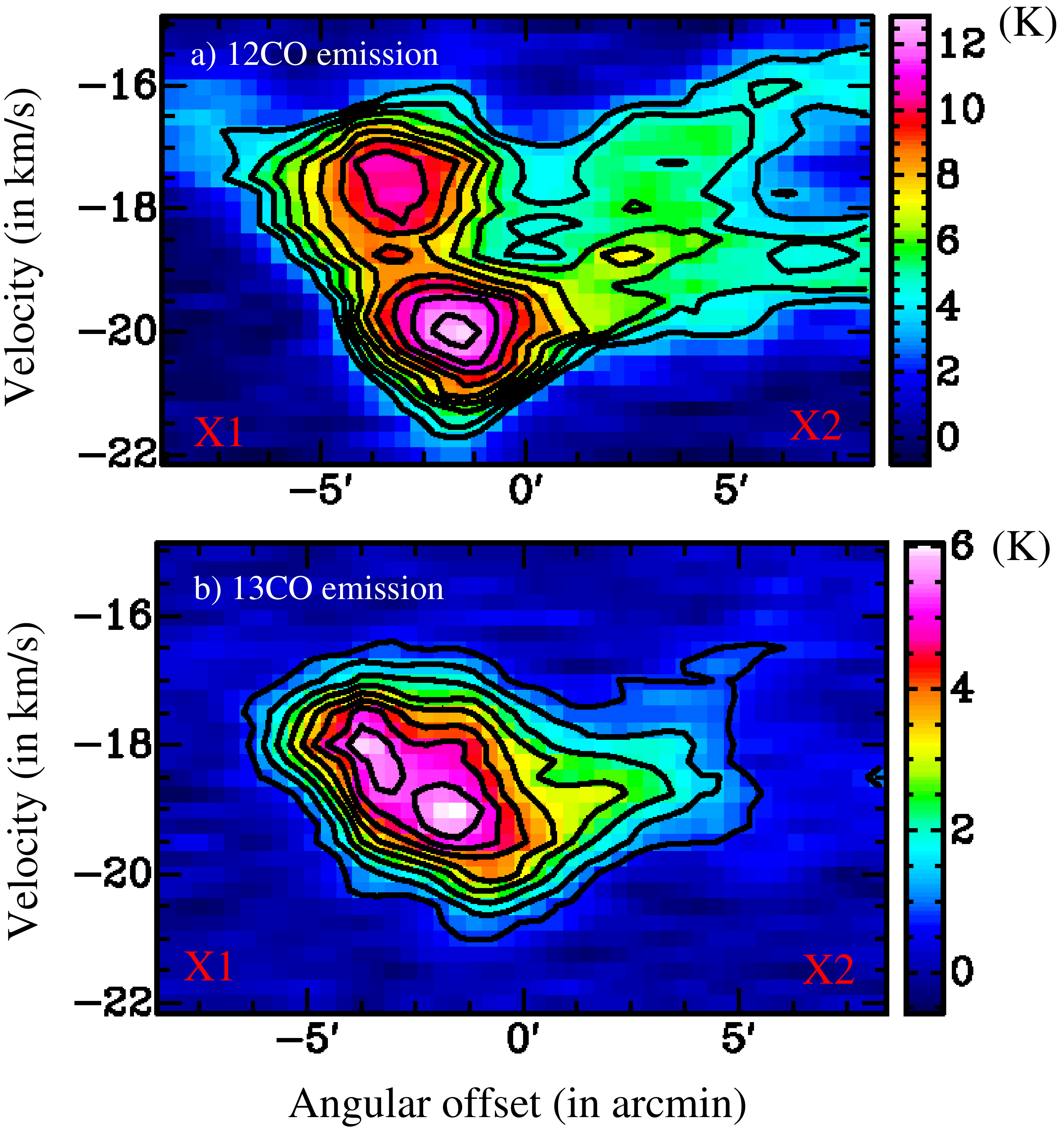}
\caption{a) A position-velocity map of $^{12}$CO along the axis (i.e., X1--X2) as highlighted in Figure~\ref{fig8}a, depicting at least two cloud components (around $-$20 and $-$17 km s$^{-1}$) along the line-of-sight. b) A position-velocity map of $^{13}$CO along the axis (i.e., X1--X2) as highlighted in Figure~\ref{fig8}a. In both the panels, contours are also plotted to show different features.}
\label{fig8v}
\end{figure*}

\begin{figure*} 
\epsscale{0.7}
\plotone{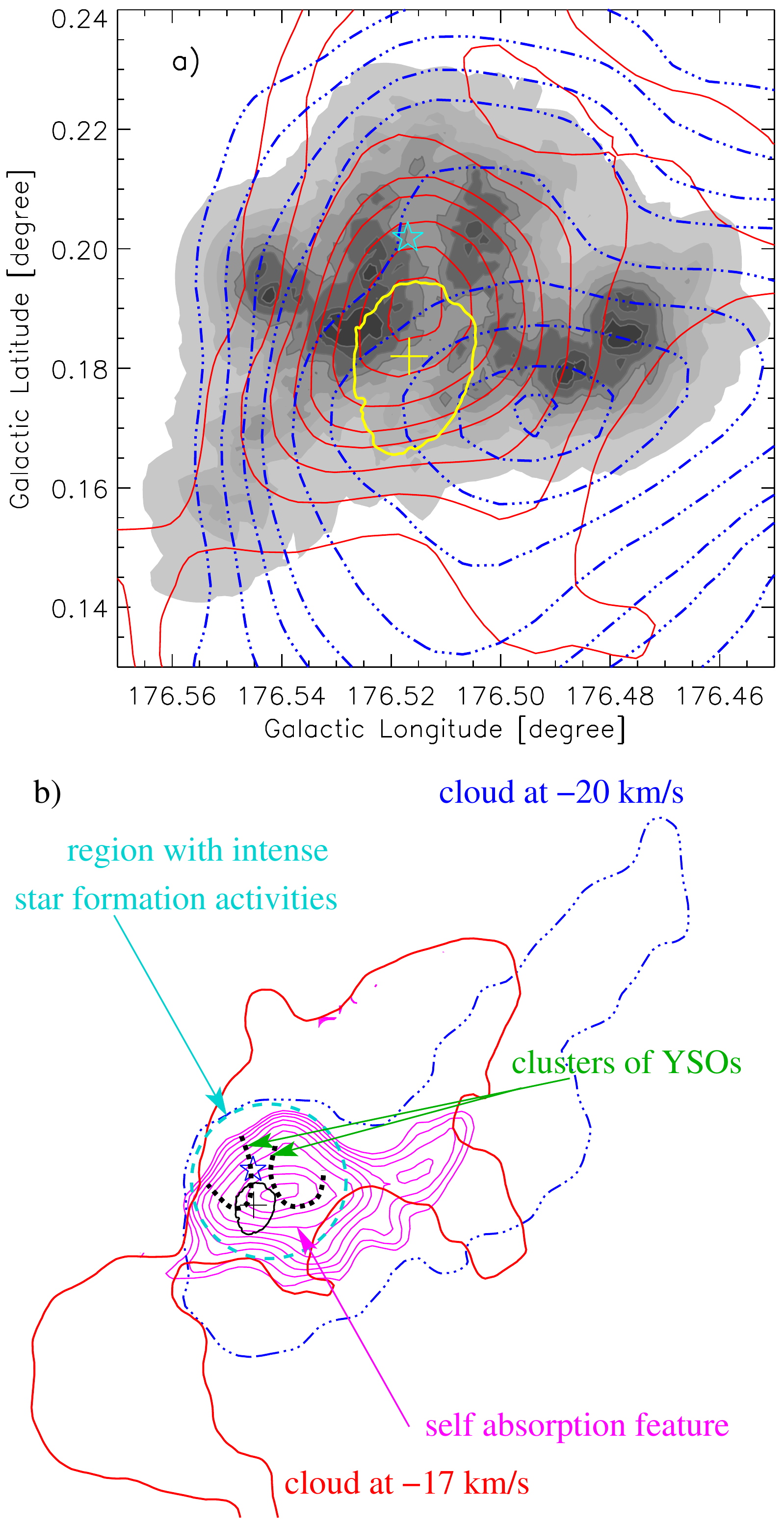}
\caption{a) Overlay of the $^{13}$CO emission contours (at [$-$17, $-$16] and [$-$20, $-$19] km s$^{-1}$) on the 
surface density contour map (see also Figure~\ref{fig11}a). The area shown in this figure is same as in Figure~\ref{fig10}c. 
The red contours show the $^{13}$CO emission at [$-$17, $-$16] km s$^{-1}$, 
while the blue contours represent the $^{13}$CO emission at [$-$20, $-$19] km s$^{-1}$ (see also Figure~\ref{fig7}d). 
Other marked symbols and the contour are similar to those shown in Figure~\ref{fig2}. 
b) A schematic figure displays the spatial distribution of two clouds (around $-$20 and $-$17 km s$^{-1}$) at large-scale (see also Figure~\ref{fig7}d). 
The region in the broken circle indicates the area associated with the intense SF activities (see also Figure~\ref{fig12}a). Two arc-like curves represent the locations of the embedded clusters of YSOs (see also Figure~\ref{fig12}a). The location of the $^{12}$CO self-absorption feature is also marked by solid 
contours (in magenta; levels =[0.8, 0.9, 1.0, 1.2, 1.4, 1.6, 1.8, 2.0, 2.2, 2.4]), which is identified based on a low ratio value ($<$ 2.5) of $^{12}$CO/$^{13}$CO at $-$18.5 km s$^{-1}$.}
\label{fig12}
\end{figure*}

Figures~\ref{fig5} and~\ref{fig6} display the integrated $^{12}$CO and $^{13}$CO velocity channel 
maps from $-$23 to $-$14 km s$^{-1}$, respectively. 
Each velocity channel map is produced by integrating the emission over 1 km s$^{-1}$ velocity interval. 
The location of the filamentary feature is also marked in both the channel maps. The velocity channel maps also suggest the presence of two velocity components (see panels at [$-$17, $-$16] and [$-$20, $-$19] km s$^{-1}$ in Figures~\ref{fig5} and~\ref{fig6}). Figures~\ref{fig7}a and~\ref{fig7}b present the first moment maps of $^{12}$CO and $^{13}$CO, respectively. The moment map is known to depict the intensity-weighted mean velocity of the emitting gas. Figure~\ref{fig7}c displays a color-composite image of AFGL 5157 with the $^{13}$CO maps 
at [$-$17, $-$16] and [$-$20, $-$19] km s$^{-1}$ in red and green, respectively. 
The filamentary feature observed in the {\it Herschel} column density map is also highlighted in Figures~\ref{fig7}a,~\ref{fig7}b, 
and~\ref{fig7}c. It seems that this color-composite image reproduces the velocity field as observed in the first moment maps of $^{12}$CO and $^{13}$CO. 
Therefore, it implies the presence of two elongated filamentary molecular clouds in the direction of 
our selected target area (see Figure~\ref{fig7}b). 
In Figure~\ref{fig7}d, we have overlaid the $^{13}$CO emission contours at [$-$17, $-$16] and [$-$20, $-$19] km s$^{-1}$ 
on the {\it Herschel} column density map, suggesting the distribution of the molecular gas 
associated with two cloud components (around $-$20 and $-$17 km s$^{-1}$) toward the column density materials. 
Interestingly, the central part of the filamentary feature is found at the common zones of the clouds 
at [$-$17, $-$16] and [$-$20, $-$19] km s$^{-1}$, illustrating the spatial connections of the 
two cloud components. Previously, the NH$_{3}$ (1, 1) emission toward the dense cores was observed around $-$17 km s$^{-1}$ \citep[see Table~1 in][]{fontani12}. 
Furthermore, in Figure~\ref{fig4}c, we have also observed an 
almost flattened profile between two velocity peaks around $-$20 and $-$17 km s$^{-1}$, suggesting the 
velocity connections of these two cloud components. 
Figure~\ref{fig8}a shows the integrated $^{12}$CO emission map, which is displayed only for the comparison (see also Figure~\ref{fig4}a). 
Figures~\ref{fig8}b and~\ref{fig8}c display the Latitude-velocity map of $^{12}$CO and Longitude-velocity 
map of $^{12}$CO, respectively. Both these position-velocity maps reveal two velocity components (around $-$20 and $-$17 km s$^{-1}$). 
In the velocity space, these two velocity peaks are also interconnected by a lower-intensity intermediate velocity emission, suggesting their connection in velocity. 
In Figures~\ref{fig8v}a and~\ref{fig8v}b, we have also obtained the position-velocity maps of $^{12}$CO and $^{13}$CO along the axis (i.e., X1--X2) 
as marked in Figure~\ref{fig8}a, respectively. Both these position-velocity maps also favour the presence of two velocity components, and their connection in velocity. 
In the direction of the central part of the filamentary feature, Figure~\ref{fig12}a displays the spatial connections of 
two cloud components of $^{13}$CO (at [$-$17, $-$16] and [$-$20, $-$19] km s$^{-1}$) against the surface density map of YSOs. 

In the large-scale area, a schematic figure is shown in Figure~\ref{fig12}b, which displays the spatial distribution of two clouds (around $-$20 and $-$17 km s$^{-1}$) 
and the area associated with the intense SF activities (see a broken circle in Figure~\ref{fig12}b). We have also highlighted the locations of the embedded clusters by 
two arc-like curves in the figure. 
The location of the $^{12}$CO self-absorption feature is also indicated by solid magenta contours in the figure.

An implication of these results is discussed in Section~\ref{sec:disc}.
\section{Discussion}
\label{sec:disc}
In this paper, we have employed the {\it Herschel} column density and temperature maps, which have provided 
a new pictorial view of AFGL 5157 (see Figure~\ref{fig2}). 
These maps have revealed at least three embedded filaments in the selected target field (see Section~\ref{subsec:u1}). 
Among these filaments, one elongated filamentary feature (having length $\sim$8.3 pc and mass $\sim$1170 M$_{\odot}$) 
is depicted, and its central part is visually seen as the junction point of the other 
filaments (see Figure~\ref{fig2}b). Embedded clusters of YSOs are depicted toward the central 
part, and are distributed around the infrared shell. Our observational results reveal that SF activities are 
concentrated mainly toward the central part of the filamentary feature (see Section~\ref{subsec:phot1}). 
The warmest region (at T$_{d}$ $\sim$20--26.5~K) in the central part of the filamentary feature 
spatially matches with the infrared shell, which hosts the diffuse H$\alpha$ emission as well as the previously known 
evolved infrared cluster. 
The {\it Herschel} column density map displays no emission toward the warmest region or the area with the diffuse H$\alpha$ emission, suggesting the impact of the massive stars in their vicinity. 
Different pressure components driven by massive stars (i.e., pressure of an H\,{\sc ii} region $(P_{HII})$, radiation pressure (P$_{rad}$), 
and stellar wind ram pressure (P$_{wind}$)) can be attributed to the feedback of massive stars \citep[e.g.,][]{bressert12,dewangan17}.
In Section~\ref{subsec:u1}, the presence of the PDRs in the central part of 
the filamentary feature is inferred, and the extension of the PDRs is depicted 
through the detection of the extended temperature structure in the {\it Herschel} temperature map. 
These results together indicate that the physical environment of AFGL 5157 appears to be influenced by massive stars located 
in the evolved infrared cluster \citep[age $\sim$10$^{6}$--10$^{7}$ yr;][]{chen03}. 
Mean ages of Class~I and Class~II YSOs have been reported to be $\sim$0.44 Myr and $\sim$1--2 Myr, respectively \citep{evans09}. 
Hence, it is likely that the expanding shell associated with the evolved infrared cluster might have influenced the birth of the youngest population in the site \citep[see also][]{chen03}. 
However, the birth process of massive stars and the infrared cluster is not yet understood in AFGL 5157.

In Section~\ref{sec:coem}, we find that the observed velocity field in the first moment maps of $^{12}$CO and $^{13}$CO can be explained by the presence of two filamentary molecular clouds (length $\sim$12.5 pc) around $-$20 and $-$17 km s$^{-1}$ in the direction of our selected target area. Therefore, the hypothesis that the star formation activity associated with AFGL 5157 involves a cloud-cloud collision should be explored. To examine SF triggered by head-on collisions, \citet{balfour17} carried out the smoothed particle 
hydrodynamics simulations and found the filamentary structures within colliding clouds. 
Previously, there are some limited examples available in the 
literature (such as, W33A \citep{galvan10}, L1641-N \citep{nakamura12}, Rosette Nebula \citep{schneider12}, Infrared dark cloud G035.39$-$00.33 \citep{henshaw13}, 
Serpens \citep{duarte11,nakamura14}, Sh 2-237 \citep{dewangan17}, and AFGL 5142 \citep{dewangan19}), 
where the collision/interaction of filaments has been proposed to explain the SF history. 
As mentioned in Section~\ref{sec:intro}, theoretically, massive stars and clusters of YSOs can be produced by the collision/interaction of two clouds \citep[e.g.,][and references therein]{loren76,habe92,inoue13,fukui14}. 
\citet{habe92} studied theoretical simulations of head-on collisions of two non-identical clouds, 
and found gravitationally unstable cores/clumps at the interface of these two clouds due to the effect of their compression.
Hence, the shock-compressed interface layer can be heated, while the molecular clouds on the trailing side will be cold \citep[e.g.,][]{loren76}. 
In this context, one may observe self-absorption of $^{12}$CO line toward the interface of cloud-cloud collisions \citep{loren76}. 
In the simulation of \citet{habe92}, a small cloud creates a cavity in the large cloud in the collision event. 
After the formation of massive OB stars, the cavity is filled with the ionized emission. 
Additionally, in the collision process, one can find the spatial and velocity connections of two molecular clouds \citep[e.g.,][]{torii17,dewangan19a}. 
One can also expect an almost flattened profile between two velocity peaks in the molecular spectrum. 
Furthermore, in the velocity space, one can also observe a connection of two molecular clouds through a feature at the intermediate velocity range, 
suggesting the presence of a bridge-like feature. The observed bridge feature may indicate the existence of a 
compressed layer of gas due to two colliding clouds/flows \citep[e.g.,][]{haworth15a,haworth15b,torii17}.
 
Based on the analysis of the $^{12}$CO and $^{13}$CO line data, a significant self-absorption of $^{12}$CO line at $-$18.5 km s$^{-1}$ is evident toward the common areas 
of two clouds (see Figure~\ref{fig12}b and also Section~\ref{sec:coem}). 
The peak contours of the self-absorption feature are seen away from the peaks of the two cloud components (i.e., $-$20 and $-$17 km s$^{-1}$). 
We also find that the peak of the $^{12}$CO emission around $-$17 km s$^{-1}$ is seen toward the {\it Herschel} clump ``A", while the peak of the $^{12}$CO emission around $-$20 km s$^{-1}$ is nearly seen in the direction of the {\it Herschel} clump ``E". 
Double peaks in the CS spectra toward two dense cores have also been observed by \citet{lee11}, which are seen toward the embedded clusters (see Section~\ref{subsec:phot1}). 
In the velocity space of $^{12}$CO, a bridge-like feature at the intermediate velocity range between two velocity peaks (i.e., $-$20 and $-$17 km s$^{-1}$) is noticed (see Figures~\ref{fig8}b and~\ref{fig8}c), which may be due to self-absorption (see Section~\ref{sec:coem}). 
However, the weaker emission connecting two velocity components is also seen in the position-velocity map of $^{13}$CO (see Figure~\ref{fig8v}b). 
Hence, there is also a signature of a bridge-like feature in $^{13}$CO, indicating the existence of the turbulent gas excited by the collision process. 
High resolution optically thin line data will be useful to further examine the bridge-like feature. 
Previously, using the $^{12}$CO spectra, \citet{torii17} also reported a similar feature in the star-forming region M20 (see their Figure~14), which is 
believed to be a site of cloud-cloud collision. 

The spatial fit of ``Keyhole/intensity-depression" and ``Key/intensity-enhancement" features is considered as an another piece of observational evidence of the collision event \citep[e.g.,][]{torii17,fukui18,dewangan19a}. In general, an intensity/gas deficient region in the molecular map is referred to as a ``Keyhole/cavity" feature, 
while an intensity-enhancement area is called as a ``Key" feature \citep[see][for more details]{fukui18}.  
Using the {\it Herschel} column density map (resolution $\sim$12$''$), a low column density area or intensity-depression or cavity is seen around the 
{\it Herschel} clumps ``A"--``D", and is filled with the diffuse H$\alpha$ emission/ionized gas (see Figure~\ref{fig2a}a). However, due to coarse beam sizes ($\sim$45$''$) of the molecular line data, our molecular maps do not permit to explore the ``Keyhole" and ``Key" features toward 
the central part of the {\it Herschel} filamentary feature. 

In AFGL 5157, the central part of the {\it Herschel} filamentary feature is observed at the overlapping zones of 
the two filamentary molecular clouds, where embedded clusters of YSOs, evolved infrared clusters, massive protostar candidates, massive stars, and massive clumps are located (see Figures~\ref{fig12}a and~\ref{fig12}b). 
In the direction of the central part, a very high value of the column density 
is found to be $\sim$3.7 $\times$ 10$^{23}$ cm$^{-2}$. Overall, the observed SF activities in AFGL 5157 are mainly depicted toward the common areas of the filamentary molecular clouds. 
Keeping in mind the predictions of the collision process, our observational findings suggest 
that the collision scenario is applicable in AFGL 5157. 
In this context, large-scale and high resolution optically thin molecular line observations will be helpful to further 
confirm the collision hypothesis in AFGL 5157. 

In order to compute the collision time-scale, we have followed the works of \citet{henshaw13} and \citet{dewangan19}.
Using the equation~1 given in \citet{dewangan19}, the time-scale of the accumulation of material at 
the collision points or the collision time-scale can be calculated with the knowledge of the collision length-scale, the observed relative velocity, 
and the ratio of the mean densities of the pre- and post-collision regions. 
It is noted that the observational determination of the exact ratio of the mean densities of the pre- and post-collision regions is not possible 
using the data utilized in this paper, and is beyond the scope of the present work. However, the increase of density in the post-collision region is expected in the collision process.
Therefore, the density ratio is likely to be higher than unity. 
We have computed the collision length-scale and the observed relative velocity to be $\sim$3.95 pc (= 2.8 pc/sin(45$\degr$)) 
and $\sim$4.25 km s$^{-1}$ (= 3.0 km s$^{-1}$/cos(45$\degr$)), respectively. 
A viewing angle of the collision is adopted to be 45$\degr$ in this calculation.
Taking into account a range of the ratios of densities (i.e., 1--10), the typical collision timescales are 
computed to be $\sim$1.8--18.6 Myr, indicating that the collision/interaction happened earlier than 1.8 Myr in AFGL 5157. 
Even though, there is uncertainty in the calculation, but the collision timescale can be 
referred to as indicative value. Considering the stellar ages discussed in this section and the collision timescale, it is possible that the collision/interaction 
of the two filamentary clouds might have triggered the birth of massive stars and YSOs in AFGL 5157. 
\section{Summary and Conclusions}
\label{sec:conc}
The present paper is directed at gaining an understanding of the formation process of massive stars and young stellar clusters 
in AFGL 5157. We have investigated an area (size $\sim$0$\degr$.43 (13.5 pc) $\times$ 0$\degr$.43 (13.5 pc)) of AFGL 5157. 
In this paper, the {\it Herschel} sub-mm continuum images and molecular line data are carefully examined to probe the physical environment of the selected site, and have allowed us to shed light on the ongoing physical process. The distribution of the embedded YSOs is studied using the photometric data at 1--8 $\mu$m. 
The observational findings of the present work are as follows:\\\\ 
$\bullet$ In the selected site, at least three embedded filaments are visually seen 
in the {\it Herschel} column density and temperature maps. Among these filaments, an elongated filamentary 
feature (having length $\sim$8.3 pc and mass $\sim$1170 M$_{\odot}$) is identified with the $N(\mathrm H_2)$ contour 
at 3.94 $\times$ 10$^{21}$ cm$^{-2}$.\\ 
$\bullet$ The central part of the filamentary feature is traced using the $N(\mathrm H_2)$ contour 
at $\sim$9--11 $\times$ 10$^{21}$ cm$^{-2}$, and contains the previously detected H$_{2}$O masers, H$_{2}$ knots, 
massive protostar candidates, and H\,{\sc ii} region.\\  
$\bullet$ The other two filaments are visually found to be directed toward the central part of the filamentary feature, and 
exhibit a temperature range of $\sim$13--13.5~K in the {\it Herschel} temperature map.\\
$\bullet$ Five massive and dense clumps (``A"--``E"; M$_{clump}$ $\sim$45--300 M$_{\odot}$; $n_{\mathrm H_2}$ $\sim$1.3--5.5 $\times$ 10$^{4}$ cm$^{-3}$) are identified toward the central part of the filamentary feature.\\
$\bullet$ The {\it Herschel} temperature map shows an extended temperature structure toward the central part of the filamentary feature. 
The infrared shell is spatially seen within this temperature structure at T$_{d}$ $\sim$20--26.5~K. 
Furthermore, an area with a substantial cold dust emission (T$_{d}$ $\sim$13.5--15~K) is also investigated inside this extended temperature structure.\\ 
$\bullet$ Based on the analysis of the {\it Spitzer} 8.0 $\mu$m image and 
the {\it Spitzer} ratio map of 4.5 $\mu$m/3.6 $\mu$m emission, the presence of PDRs in AFGL 5157 is traced. 
The observed extended temperature structure suggests the distribution of the warm dust emission 
in the PDRs, revealing the signatures of the impact of massive stars in AFGL 5157.\\
$\bullet$ The {\it Spitzer} ratio map reveals the signatures of outflow activities toward the IRDCs seen in the {\it Spitzer} 8.0 $\mu$m image, where 
the {\it Herschel} column density map traces higher column density materials. \\ 
$\bullet$ Using the $^{12}$CO and $^{13}$CO line data, the molecular gas associated with AFGL 5157 is studied 
in a velocity range of [$-22$, $-$15] km s$^{-1}$, and shows two velocity components around $-$20 and $-$17 km s$^{-1}$. 
Using the molecular line data, a bridge-like feature is investigated at the intermediate velocity range between these two velocity peaks.\\ 
$\bullet$ The central part of the {\it Herschel} filamentary feature is found at the overlapping zones of two filamentary 
molecular clouds (length $\sim$12.5 pc) around $-$20 and $-$17 km s$^{-1}$, which are also connected in the velocity space. 
Considering the observed low ratio value ($<$ 2.5) of $^{12}$CO/$^{13}$CO at $-$18.5 km s$^{-1}$, 
self-absorption of $^{12}$CO line is found toward the common areas of two clouds.\\ 
$\bullet$ Using the photometric analysis of point-like sources, embedded clusters of YSOs are found mainly toward the central part of the {\it Herschel} filamentary feature, 
and are distributed around the infrared shell. These embedded clusters are also located toward the spatial overlapping zones of the two filamentary
 molecular clouds (around $-$20 and $-$17 km s$^{-1}$).  In the direction of the central part, the maximum value of the column density 
is found to be $\sim$3.7 $\times$ 10$^{23}$ cm$^{-2}$, favouring the appropriate condition for the birth of massive stars.\\

Overall, our findings suggest that the cloud-cloud collision operated in the past, which might have triggered 
the observed SF activities in AFGL 5157.
\acknowledgments  
We thank the anonymous reviewer for several useful comments and suggestions, which greatly improved the scientific 
contents of the paper. The research work at Physical Research Laboratory is funded by the Department of Space, Government of India. 
This work is based on data obtained as part of the UKIRT Infrared Deep Sky Survey. This publication 
made use of data products from the Two Micron All Sky Survey (a joint project of the University of Massachusetts and 
the Infrared Processing and Analysis Center / California Institute of Technology, funded by NASA and NSF), archival 
data obtained with the {\it Spitzer} Space Telescope (operated by the Jet Propulsion Laboratory, California Institute 
of Technology under a contract with NASA). 
This paper makes use of data obtained as part of the INT Photometric H$\alpha$ Survey of the Northern Galactic 
Plane (IPHAS, www.iphas.org) carried out at the Isaac Newton Telescope (INT). The INT is operated on the 
island of La Palma by the Isaac Newton Group in the Spanish Observatorio del Roque de los Muchachos of 
the Instituto de Astrofisica de Canarias. The IPHAS data are processed by the Cambridge Astronomical Survey 
Unit, at the Institute of Astronomy in Cambridge. 
WISE is a joint project of the University of California and the JPL, Caltech, funded by the NASA. 
IRAF is distributed by the National Optical Astronomy Observatory, USA. The Canadian Galactic Plane Survey (CGPS) is a Canadian project with international partners. 
The Dominion Radio Astrophysical Observatory is operated as a national facility by the 
National Research Council of Canada. The Five College Radio Astronomy Observatory 
CO Survey of the Outer Galaxy was supported by NSF grant AST 94-20159. The CGPS is 
supported by a grant from the Natural Sciences and Engineering Research Council of Canada. 
%

 \begin{table*}
  \tiny
\setlength{\tabcolsep}{0.05in}
\centering
\caption{Multi-wavelength surveys used in this observational work.}
\label{utab1a}
\begin{tabular}{lcccr}
\hline 
  Survey  &  Wavelength(s)       &  Resolution ($\arcsec$)        &  Reference \\   
\hline
\hline 
Extended Outer Galaxy Survey (E-OGS)                                                                  & 2.6--2.7 mm (CO(1-0)) & $\sim$45--46        &\citet{brunt04}\\
SCUBA Legacy Catalogues and continuum maps                             &850 $\mu$m                     & $\sim$19         &\citet{difrancesco08}\\
{\it Herschel} Infrared Galactic Plane Survey (Hi-GAL)                              &70, 160, 250, 350, 500 $\mu$m                     & $\sim$5.8, $\sim$12, $\sim$18, $\sim$25, $\sim$37         &\citet{molinari10}\\
Wide Field Infrared Survey Explorer (WISE) & 12 $\mu$m                   & $\sim$6           &\citet{wright10}\\ 
Warm-{\it Spitzer} GLIMPSE360 Survey      &3.6, 4.5 $\mu$m                   & $\sim$2, $\sim$2           &\citet{whitney11,benjamin03}\\
UKIRT near-infrared Galactic Plane Survey (GPS)                                                 &1.25--2.2 $\mu$m                   &$\sim$0.8           &\citet{lawrence07}\\ 
Two Micron All Sky Survey (2MASS)                                                 &1.25--2.2 $\mu$m                  & $\sim$2.5          &\citet{skrutskie06}\\
INT Photometric H$\alpha$ Survey of the Northern Galactic Plane (IPHAS)                                                 &0.6563 $\mu$m                   &$\sim$1           &\citet{drew05}\\
\hline          
\end{tabular}
\end{table*}
 \begin{table*}
\setlength{\tabcolsep}{0.4in}
\centering
\caption{Properties of the selected {\it Herschel} clumps in AFGL 5157 (see Figures~\ref{fig2}c and~\ref{fig2a}b). 
Clump IDs, positions, deconvolved effective radius (R$_{clump}$), clump mass (M$_{clump}$), and 
average volume density ($n_{\mathrm H_2}$) are listed in the table. 
Twenty three clumps (i.e., 1--23) are labeled in Figure~\ref{fig2}c, while five clumps (i.e., A--E) are labeled in 
Figure~\ref{fig2a}b.}
\label{tab1}
\begin{tabular}{lccccccr}
\hline 
  ID  & {\it l}       &  {\it b} &R$_{clump}$       & M$_{clump}$  & $n_{\mathrm H_2}$\\   
    &  (degree)      &  (degree) &(pc)       &($M_\odot$) &(cm$^{-3}$)\\   
\hline
\hline 
  1 &	176.362 &     0.320 &	  0.29  &  25  &     3540\\
  2 &	176.457 &     0.284 &	  0.43  &  65  &     2825\\
  3 &	176.457 &     0.269 &	  0.23  &  17  &     4830\\ 
  4 &	176.474 &     0.259 &	  0.28  &  27  &     4250\\ 
  5 &	176.460 &     0.222 &	  0.33  &  35  &     3365\\        
  6 &	176.410 &     0.194 &	  0.23  &  16  &     4545\\
  7 &	176.439 &     0.229 &	  0.21  &  14  &     5225\\
  8 &	176.407 &     0.234 &	  0.25  &  20  &     4425\\ 
  9 &	176.432 &     0.215 &	  0.28  &  28  &     4410\\ 
 10 &	176.422 &     0.210 &	  0.26  &  23  &     4520\\        
 11 &	176.435 &     0.202 &	  0.25  &  20  &     4425\\
 12 &	176.465 &     0.205 &	  0.26  &  20  &     3930\\
 13 &	176.459 &     0.200 &	  0.32  &  48  &     5065\\ 
 14 &	176.444 &     0.184 &	  0.23  &  17  &     4830\\ 
 15 &	176.459 &     0.177 &	  0.35  &  48  &     3870\\        
 16 &	176.519 &     0.199 &	  0.80  & 724  &     4885\\
 17 &	176.544 &     0.172 &	  0.34  &  38  &     3340\\
 18 &	176.550 &     0.159 &	  0.26  &  28  &     5505\\ 
 19 &	176.567 &     0.149 &	  0.27  &  30  &     5270\\ 
 20 &	176.575 &     0.134 &	  0.28  &  25  &     3935\\        
 21 &	176.594 &     0.115 &	  0.35  &  45  &     3625\\
 22 &	176.609 &     0.124 &	  0.37  &  43  &     2935\\
 23 &	176.604 &     0.080 &	  0.46  &  64  &     2270\\ 
  A &   176.519 &     0.199 &     0.32	& 300  &    31650\\
  B &   176.527 &     0.195 &     0.16	&  65  &    54860\\
  C &   176.527 &     0.187 &     0.27	& 110  &    19320\\ 
  D &   176.510 &     0.179 &     0.23	&  45  &    12785\\ 
  E &   176.490 &     0.179 &     0.25	&  65  &    14380\\        
\hline          
\end{tabular}
\end{table*}

 \begin{table*}
  \tiny
\setlength{\tabcolsep}{0.05in}
\centering
\caption{Near-infrared and {\it Spitzer} IRAC/GLIMPSE photometric magnitudes of selected YSOs (see Figures~\ref{fig9}c and~\ref{fig10}c).}
\label{tab3}
\begin{tabular}{lcccccccccccccccr}
\hline 
  ID  &  {\it l}       & {\it b} &J (mag)  & H (mag) & K (mag)  &  3.6 $\mu$m (mag)       &  4.5 $\mu$m (mag) & 5.8 $\mu$m (mag)   &   8.0 $\mu$m (mag)    &  Class\\   
\hline
\hline 
   1 & 176.612 &   0.110  &    ---  &  15.75  &  14.37 &   12.72  &  11.85  &	 ---  &    ---  &  I   \\
   2 & 176.330 &   0.193  &  15.62  &  14.65  &  13.96 &   13.14  &  12.48  &	 ---  &    ---  &  I   \\
   3 & 176.441 &   0.231  &  16.00  &  14.56  &  13.53 &   12.33  &  11.57  &	 ---  &    ---  &  I   \\
   4 & 176.489 &   0.179  &    ---  &  15.92  &  13.85 &   12.46  &  11.26  &	 ---  &    ---  &  I   \\
   5 & 176.504 &   0.197  &    ---  &  13.46  &  12.74 &   11.52  &  10.71  &	 ---  &    ---  &  I   \\
   6 & 176.526 &   0.186  &  15.26  &  13.69  &  12.75 &   11.03  &  10.38  &	 ---  &    ---  &  I   \\
   7 & 176.495 &   0.188  &  17.39  &  15.77  &  15.08 &   13.68  &  13.00  &	 ---  &    ---  &  I   \\
   8 & 176.547 &   0.160  &  15.11  &  14.03  &  13.53 &   12.28  &  11.57  &	 ---  &    ---  &  I   \\
   9 & 176.542 &   0.194  &    ---  &  15.63  &  14.86 &   12.77  &  12.19  &	 ---  &    ---  &  I   \\
  10 & 176.625 &   0.056  &  16.97  &  15.69  &  14.94 &   13.64  &  13.22  &	 ---  &    ---  &  II  \\
  11 & 176.625 &   0.028  &  14.68  &  13.69  &  13.25 &   12.52  &  12.06  &	 ---  &    ---  &  II  \\
  12 & 176.633 &   0.092  &  15.10  &  14.31  &  13.81 &   13.15  &  12.85  &	 ---  &    ---  &  II  \\
  13 & 176.354 &   0.422  &  15.27  &  14.69  &  14.68 &   14.49  &  14.35  &	 ---  &    ---  &  II  \\
  14 & 176.377 &   0.387  &  15.17  &  14.69  &  14.71 &   14.51  &  14.38  &	 ---  &    ---  &  II  \\
  15 & 176.334 &   0.399  &  15.67  &  14.82  &  14.41 &   14.43  &  14.07  &	 ---  &    ---  &  II  \\
  16 & 176.356 &   0.191  &  15.31  &  14.79  &  14.75 &   14.39  &  14.26  &	 ---  &    ---  &  II  \\
  17 & 176.565 &   0.294  &  13.92  &  13.08  &  12.49 &   11.55  &  11.16  &	 ---  &    ---  &  II  \\
  18 & 176.445 &   0.338  &  16.51  &  15.59  &  14.99 &   14.56  &  14.26  &	 ---  &    ---  &  II  \\
  19 & 176.460 &   0.189  &  15.83  &  14.64  &  14.14 &   13.28  &  13.07  &	 ---  &    ---  &  II  \\
  20 & 176.474 &   0.198  &  14.97  &  14.03  &  13.47 &   12.59  &  12.26  &	 ---  &    ---  &  II  \\
  21 & 176.461 &   0.200  &  13.51  &  12.25  &  11.33 &   10.20  &   9.69  &	 ---  &    ---  &  II  \\
  22 & 176.510 &   0.232  &  15.27  &  14.44  &  14.16 &   13.81  &  13.59  &	 ---  &    ---  &  II  \\
  23 & 176.455 &   0.274  &  12.75  &  11.81  &  11.21 &   10.18  &   9.89  &	 ---  &    ---  &  II  \\
  24 & 176.455 &   0.256  &  14.44  &  13.39  &  12.96 &   12.45  &  12.19  &	 ---  &    ---  &  II  \\
  25 & 176.481 &   0.221  &    ---  &  16.12  &  15.26 &   13.98  &  13.63  &	 ---  &    ---  &  II  \\
  26 & 176.468 &   0.292  &  15.47  &  14.94  &  14.69 &   14.65  &  14.30  &	 ---  &    ---  &  II  \\
  27 & 176.464 &   0.284  &    ---  &  14.86  &  13.69 &   12.44  &  11.78  &	 ---  &    ---  &  II  \\
  28 & 176.422 &   0.200  &  17.24  &  15.87  &  15.29 &   14.31  &  13.95  &	 ---  &    ---  &  II  \\
  29 & 176.454 &   0.128  &  10.88  &  10.52  &  10.25 &    9.87  &   9.58  &	 ---  &    ---  &  II  \\
  30 & 176.583 &   0.252  &  15.83  &  14.83  &  14.51 &   13.97  &  13.68  &	 ---  &    ---  &  II  \\
  31 & 176.486 &   0.177  &  15.26  &  14.31  &  13.66 &   12.78  &  12.20  &	 ---  &    ---  &  II  \\
  32 & 176.479 &   0.168  &  15.26  &  13.95  &  13.10 &   11.57  &  11.04  &	 ---  &    ---  &  II  \\
  33 & 176.520 &   0.153  &  16.31  &  15.32  &  14.77 &   13.93  &  13.42  &	 ---  &    ---  &  II  \\
  34 & 176.492 &   0.206  &  16.38  &  14.66  &  13.75 &   12.58  &  12.13  &	 ---  &    ---  &  II  \\
  35 & 176.503 &   0.203  &  15.88  &  14.90  &  14.48 &   14.18  &  13.91  &	 ---  &    ---  &  II  \\
  36 & 176.506 &   0.207  &  11.89  &  11.01  &  10.36 &    9.38  &   8.90  &	 ---  &    ---  &  II  \\
  37 & 176.526 &   0.194  &  16.05  &  14.92  &  14.23 &   13.26  &  12.78  &	 ---  &    ---  &  II  \\
  38 & 176.525 &   0.186  &  15.80  &  13.82  &  12.64 &   11.50  &  11.05  &	 ---  &    ---  &  II  \\
  39 & 176.525 &   0.195  &  15.86  &  13.60  &  12.21 &   10.89  &  10.24  &	 ---  &    ---  &  II  \\
  40 & 176.511 &   0.195  &    ---  &  14.99  &  13.37 &   11.44  &  10.77  &	 ---  &    ---  &  II  \\
  41 & 176.512 &   0.186  &  15.97  &  14.48  &  13.45 &   12.02  &  11.41  &	 ---  &    ---  &  II  \\
  42 & 176.513 &   0.169  &    ---  &  15.89  &  15.37 &   14.40  &  14.11  &	 ---  &    ---  &  II  \\
  43 & 176.515 &   0.179  &    ---  &  11.04  &  10.24 &    8.86  &   8.36  &	 ---  &    ---  &  II  \\
  44 & 176.497 &   0.177  &  15.68  &  14.13  &  13.27 &   12.02  &  11.47  &	 ---  &    ---  &  II  \\
  45 & 176.506 &   0.170  &  16.08  &  14.99  &  14.72 &   14.19  &  13.99  &	 ---  &    ---  &  II  \\
  46 & 176.505 &   0.182  &  16.92  &  15.37  &  14.48 &   13.20  &  12.71  &	 ---  &    ---  &  II  \\
  47 & 176.545 &   0.123  &  15.46  &  14.47  &  14.34 &   13.87  &  13.74  &	 ---  &    ---  &  II  \\
  48 & 176.550 &   0.134  &    ---  &  16.07  &  15.34 &   14.70  &  14.07  &	 ---  &    ---  &  II  \\
  49 & 176.569 &   0.146  &  16.43  &  15.94  &  15.27 &   14.40  &  13.80  &	 ---  &    ---  &  II  \\
  50 & 176.585 &   0.146  &  14.72  &  13.79  &  13.50 &   13.02  &  12.81  &	 ---  &    ---  &  II  \\
  51 & 176.577 &   0.112  &  15.08  &  14.44  &  14.19 &   13.74  &  13.56  &	 ---  &    ---  &  II  \\
  52 & 176.588 &   0.130  &  14.85  &  13.93  &  13.53 &   12.63  &  12.19  &	 ---  &    ---  &  II  \\
  53 & 176.583 &   0.131  &  15.80  &  14.74  &  14.29 &   13.85  &  13.57  &	 ---  &    ---  &  II  \\
  54 & 176.551 &   0.167  &  17.30  &  16.08  &  15.44 &   14.36  &  13.83  &	 ---  &    ---  &  II  \\
  55 & 176.580 &   0.177  &  16.36  &  15.54  &  15.22 &   14.80  &  14.56  &	 ---  &    ---  &  II  \\
  56 & 176.533 &   0.188  &    ---  &  16.32  &  15.14 &   13.99  &  13.45  &	 ---  &    ---  &  II  \\
  57 & 176.538 &   0.200  &  16.80  &  15.81  &  15.13 &   14.14  &  13.72  &	 ---  &    ---  &  II  \\
  58 & 176.530 &   0.205  &  16.76  &  15.46  &  14.64 &   14.16  &  13.74  &	 ---  &    ---  &  II  \\
  59 & 176.547 &   0.199  &  15.13  &  13.99  &  13.45 &   12.20  &  11.82  &	 ---  &    ---  &  II  \\
  60 & 176.535 &   0.189  &  16.94  &  15.66  &  14.86 &   14.41  &  14.00  &	 ---  &    ---  &  II  \\
  61 & 176.544 &   0.177  &  16.76  &  15.85  &  15.48 &   14.39  &  13.89  &	 ---  &    ---  &  II  \\
  62 & 176.541 &   0.195  &    ---  &  14.36  &  13.86 &   13.11  &  12.73  &	 ---  &    ---  &  II  \\
  63 & 176.545 &   0.194  &  16.41  &  15.61  &  15.19 &   14.46  &  14.07  &	 ---  &    ---  &  II  \\
  64 & 176.541 &   0.189  &  17.04  &  16.01  &  15.09 &   14.04  &  13.52  &	 ---  &    ---  &  II  \\
  65 & 176.540 &   0.194  &  15.09  &  13.97  &  13.38 &   12.44  &  12.07  &	 ---  &    ---  &  II  \\
  66 & 176.546 &   0.193  &  16.80  &  15.77  &  15.34 &   14.77  &  14.45  &	 ---  &    ---  &  II  \\
  67 & 176.472 &   0.130  &  14.99  &  13.82  &  13.00 &   12.08  &  11.61  &	 ---  &    ---  &  II  \\
  68 & 176.515 &   0.114  &  17.05  &  15.98  &  15.60 &   14.47  &  14.23  &	 ---  &    ---  &  II  \\
  69 & 176.265 &   0.393  &  13.40  &  12.71  &  12.08 &   11.15  &  10.76  &	 ---  &    ---  &  II  \\
  70 & 176.289 &   0.079  &  14.26  &  13.68  &  13.62 &   13.82  &  13.57  &	 ---  &    ---  &  II  \\
  71 & 176.274 &   0.161  &  15.73  &  15.14  &  14.89 &   14.78  &  14.52  &	 ---  &    ---  &  II  \\
  72 & 176.487 &   0.181  &    ---  &	 ---  &    --- &   15.43  &  14.40  &  13.44  &  13.19  &  II  \\
  73 & 176.485 &   0.178  &    ---  &	 ---  &    --- &   13.61  &  12.69  &  11.71  &  10.57  &  II  \\
  74 & 176.491 &   0.181  &    ---  &	 ---  &    --- &   16.87  &  14.59  &  13.65  &  12.31  &  II  \\
  75 & 176.523 &   0.199  &    ---  &	 ---  &    --- &   12.87  &  11.70  &  10.95  &  10.21  &  II  \\    
  76 & 176.494 &   0.184  &    ---  &	 ---  &    --- &   11.51  &   9.95  &	9.12  &   8.34  &  II  \\
  77 & 176.528 &   0.188  &    ---  &	 ---  &    --- &   13.14  &  12.18  &  11.38  &  10.47  &  II  \\
  78 & 176.529 &   0.188  &    ---  &	 ---  &    --- &   13.46  &  12.34  &  11.28  &  10.37  &  II  \\
  79 & 176.533 &   0.179  &    ---  &	 ---  &    --- &   15.40  &  14.64  &  13.55  &  12.56  &  II  \\
  80 & 176.485 &   0.194  &    ---  &	 ---  &    --- &   15.99  &  15.23  &  14.76  &  13.01  &  II  \\
  81 & 176.517 &   0.224  &    ---  &	 ---  &    --- &   14.73  &  14.19  &  13.38  &  12.30  &  II  \\
  82 & 176.501 &   0.201  &    ---  &	 ---  &    --- &   14.40  &  13.92  &  13.40  &  11.97  &  II  \\
  83 & 176.510 &   0.201  &    ---  &	 ---  &    --- &   12.62  &  11.17  &  10.52  &   9.47  &  II  \\
\hline          
\end{tabular}
\end{table*}

 \begin{table*}
  \tiny
\setlength{\tabcolsep}{0.05in}
\centering
\caption{Continuation of Table~3}
\begin{tabular}{lcccccccccccccccr}
\hline 
  ID  &  {\it l}       & {\it b} &J (mag)  & H (mag) & K (mag)  &  3.6 $\mu$m (mag)       &  4.5 $\mu$m (mag) & 5.8 $\mu$m (mag)   &   8.0 $\mu$m (mag)    &  Class\\   
\hline
\hline 
  84 & 176.490 &   0.202  &    ---  &	 ---  &    --- &   15.15  &  14.72  &  14.49  &  12.34  & I    \\
  85 & 176.496 &   0.187  &    ---  &	 ---  &    --- &   15.78  &  15.29  &  14.64  &  13.44  & I    \\
  86 & 176.465 &   0.195  &    ---  &	 ---  &    --- &   15.63  &  15.17  &  14.66  &  14.03  &  II  \\
  87 & 176.487 &   0.178  &    ---  &	 ---  &    --- &   12.57  &  11.86  &  11.38  &  10.62  &  II  \\
  88 & 176.476 &   0.182  &    ---  &	 ---  &    --- &   14.35  &  13.69  &  13.02  &  12.27  &  II  \\
  89 & 176.518 &   0.202  &    ---  &	 ---  &    --- &   13.10  &  12.28  &  11.77  &  11.07  &  II  \\
  90 & 176.495 &   0.184  &    ---  &	 ---  &    --- &   12.64  &  12.24  &  11.82  &  11.20  &  II  \\
  91 & 176.526 &   0.210  &    ---  &	 ---  &    --- &   14.64  &  14.12  &  13.80  &  12.75  &  II  \\
  92 & 176.540 &   0.179  &    ---  &	 ---  &    --- &   15.47  &  14.71  &  14.30  &  13.20  &  II  \\
  93 & 176.554 &   0.192  &    ---  &	 ---  &    --- &   14.60  &  14.17  &  13.95  &  13.19  &  II  \\
  94 & 176.437 &   0.230  &  18.64  &  16.73  &  15.46 &     ---  &    ---  &	 ---  &    ---  &  II  \\
  95 & 176.462 &   0.195  &  17.08  &  15.27  &  14.10 &     ---  &    ---  &	 ---  &    ---  &  II  \\
  96 & 176.456 &   0.287  &  19.63  &  17.86  &  16.78 &     ---  &    ---  &	 ---  &    ---  &  II  \\
  97 & 176.542 &   0.177  &  16.21  &  14.25  &  12.92 &     ---  &    ---  &	 ---  &    ---  &  II  \\
  98 & 176.546 &   0.163  &  20.17  &  17.29  &  15.80 &     ---  &    ---  &	 ---  &    ---  &  II  \\
  99 & 176.546 &   0.149  &  19.18  &  17.49  &  16.29 &     ---  &    ---  &	 ---  &    ---  &  II  \\
 100 & 176.542 &   0.170  &    ---  &  18.47  &  16.76 &     ---  &    ---  &	 ---  &    ---  &  II  \\
 101 & 176.437 &   0.201  &  19.56  &  17.59  &  16.52 &     ---  &    ---  &	 ---  &    ---  &  II  \\
 102 & 176.501 &   0.211  &  19.83  &  16.81  &  15.11 &     ---  &    ---  &	 ---  &    ---  &  II  \\
 103 & 176.501 &   0.206  &  17.25  &  14.19  &  12.42 &     ---  &    ---  &	 ---  &    ---  &  II  \\
 104 & 176.505 &   0.207  &  18.28  &  16.11  &  14.30 &     ---  &    ---  &	 ---  &    ---  &  II  \\
 105 & 176.503 &   0.197  &  18.28  &  16.91  &  15.81 &     ---  &    ---  &	 ---  &    ---  &  II  \\
 106 & 176.503 &   0.191  &  20.22  &  17.63  &  16.32 &     ---  &    ---  &	 ---  &    ---  &  II  \\
 107 & 176.505 &   0.190  &  20.25  &  17.51  &  15.98 &     ---  &    ---  &	 ---  &    ---  &  II  \\
 108 & 176.502 &   0.182  &  18.54  &  16.42  &  15.33 &     ---  &    ---  &	 ---  &    ---  &  II  \\
 109 & 176.502 &   0.181  &  19.15  &  17.41  &  16.09 &     ---  &    ---  &	 ---  &    ---  &  II  \\
 110 & 176.505 &   0.179  &  19.73  &  16.86  &  14.79 &     ---  &    ---  &	 ---  &    ---  &  II  \\
 111 & 176.506 &   0.200  &    ---  &  17.28  &  15.27 &     ---  &    ---  &	 ---  &    ---  &  II  \\
 112 & 176.521 &   0.213  &  19.82  &  16.88  &  15.37 &     ---  &    ---  &	 ---  &    ---  &  II  \\
 113 & 176.524 &   0.204  &  18.51  &  16.05  &  14.70 &     ---  &    ---  &	 ---  &    ---  &  II  \\
 114 & 176.523 &   0.189  &  19.43  &  16.37  &  14.47 &     ---  &    ---  &	 ---  &    ---  &  II  \\
 115 & 176.521 &   0.189  &  14.54  &  12.92  &  11.71 &     ---  &    ---  &	 ---  &    ---  &  II  \\
 116 & 176.524 &   0.186  &  18.43  &  15.53  &  13.40 &     ---  &    ---  &	 ---  &    ---  &  II  \\
 117 & 176.527 &   0.167  &  18.70  &  16.67  &  15.50 &     ---  &    ---  &	 ---  &    ---  &  II  \\
 118 & 176.525 &   0.205  &    ---  &  18.27  &  17.17 &     ---  &    ---  &	 ---  &    ---  &  II  \\
 119 & 176.521 &   0.194  &    ---  &  18.29  &  16.72 &     ---  &    ---  &	 ---  &    ---  &  II  \\
 120 & 176.522 &   0.196  &    ---  &  17.65  &  15.99 &     ---  &    ---  &	 ---  &    ---  &  II  \\
 121 & 176.526 &   0.198  &    ---  &  18.57  &  15.81 &     ---  &    ---  &	 ---  &    ---  &  II  \\
 122 & 176.485 &   0.248  &  18.45  &  16.44  &  15.07 &     ---  &    ---  &	 ---  &    ---  &  II  \\
 123 & 176.482 &   0.183  &  20.41  &  18.21  &  17.05 &     ---  &    ---  &	 ---  &    ---  &  II  \\
 124 & 176.483 &   0.186  &    ---  &  18.54  &  17.22 &     ---  &    ---  &	 ---  &    ---  &  II  \\
 125 & 176.533 &   0.191  &  19.46  &  17.91  &  16.69 &     ---  &    ---  &	 ---  &    ---  &  II  \\
 126 & 176.529 &   0.185  &  19.44  &  16.77  &  15.26 &     ---  &    ---  &	 ---  &    ---  &  II  \\
 127 & 176.529 &   0.184  &  18.97  &  16.92  &  15.62 &     ---  &    ---  &	 ---  &    ---  &  II  \\
 128 & 176.533 &   0.183  &    ---  &  17.63  &  15.87 &     ---  &    ---  &	 ---  &    ---  &  II  \\
 129 & 176.528 &   0.175  &    ---  &  17.99  &  16.31 &     ---  &    ---  &	 ---  &    ---  &  II  \\
 130 & 176.535 &   0.195  &  18.35  &  16.61  &  15.60 &     ---  &    ---  &	 ---  &    ---  &  II  \\
 131 & 176.539 &   0.165  &  18.15  &  16.09  &  14.95 &     ---  &    ---  &	 ---  &    ---  &  II  \\
 132 & 176.476 &   0.189  &  19.44  &  17.15  &  15.98 &     ---  &    ---  &	 ---  &    ---  &  II  \\
 133 & 176.477 &   0.187  &  19.59  &  17.45  &  16.16 &     ---  &    ---  &	 ---  &    ---  &  II  \\
 134 & 176.477 &   0.188  &  15.11  &  13.32  &  12.16 &     ---  &    ---  &	 ---  &    ---  &  II  \\
 135 & 176.478 &   0.187  &  16.47  &  14.75  &  13.65 &     ---  &    ---  &	 ---  &    ---  &  II  \\
 136 & 176.478 &   0.188  &  15.87  &  13.56  &  12.11 &     ---  &    ---  &	 ---  &    ---  &  II  \\
 137 & 176.474 &   0.178  &  18.64  &  16.53  &  15.20 &     ---  &    ---  &	 ---  &    ---  &  II  \\
 138 & 176.478 &   0.182  &  19.99  &  16.99  &  15.32 &     ---  &    ---  &	 ---  &    ---  &  II  \\
 139 & 176.474 &   0.180  &    ---  &  17.74  &  15.82 &     ---  &    ---  &	 ---  &    ---  &  II  \\
 140 & 176.477 &   0.183  &    ---  &  18.22  &  16.69 &     ---  &    ---  &	 ---  &    ---  &  II  \\
 141 & 176.468 &   0.190  &  19.22  &  17.53  &  16.40 &     ---  &    ---  &	 ---  &    ---  &  II  \\
 142 & 176.466 &   0.178  &    ---  &  18.33  &  17.28 &     ---  &    ---  &	 ---  &    ---  &  II  \\
 143 & 176.646 &   0.020  &  18.52  &  16.69  &  15.18 &     ---  &    ---  &	 ---  &    ---  &  II  \\
 144 & 176.514 &   0.219  &  19.28  &  18.64  &  17.54 &     ---  &    ---  &	 ---  &    ---  &  II  \\
 145 & 176.518 &   0.219  &  18.95  &  16.83  &  15.59 &     ---  &    ---  &	 ---  &    ---  &  II  \\
 146 & 176.517 &   0.215  &  19.34  &  17.94  &  16.93 &     ---  &    ---  &	 ---  &    ---  &  II  \\
 147 & 176.521 &   0.203  &  19.31  &  17.11  &  15.68 &     ---  &    ---  &	 ---  &    ---  &  II  \\
 148 & 176.515 &   0.192  &  16.50  &  14.53  &  13.31 &     ---  &    ---  &	 ---  &    ---  &  II  \\
 149 & 176.517 &   0.189  &  17.09  &  15.39  &  14.27 &     ---  &    ---  &	 ---  &    ---  &  II  \\
 150 & 176.516 &   0.159  &  19.53  &  18.54  &  17.50 &     ---  &    ---  &	 ---  &    ---  &  II  \\
 151 & 176.271 &   0.398  &  19.96  &  18.61  &  17.54 &     ---  &    ---  &	 ---  &    ---  &  II  \\
 152 & 176.452 &   0.263  &  19.86  &  18.31  &  17.07 &     ---  &    ---  &	 ---  &    ---  &  II  \\
 153 & 176.553 &   0.150  &  19.07  &  17.72  &  16.59 &     ---  &    ---  &	 ---  &    ---  &  II  \\
 154 & 176.510 &   0.213  &  20.36  &  18.15  &  16.93 &     ---  &    ---  &	 ---  &    ---  &  II  \\
 155 & 176.510 &   0.211  &  19.90  &  17.09  &  15.46 &     ---  &    ---  &	 ---  &    ---  &  II  \\
 156 & 176.513 &   0.199  &  19.45  &  18.67  &  17.60 &     ---  &    ---  &	 ---  &    ---  &  II  \\
 157 & 176.510 &   0.178  &  17.55  &  15.29  &  14.01 &     ---  &    ---  &	 ---  &    ---  &  II  \\
 158 & 176.511 &   0.170  &  19.93  &  18.38  &  17.08 &     ---  &    ---  &	 ---  &    ---  &  II  \\
 159 & 176.491 &   0.215  &  19.77  &  17.72  &  16.40 &     ---  &    ---  &	 ---  &    ---  &  II  \\
 160 & 176.492 &   0.215  &  18.87  &  16.78  &  15.63 &     ---  &    ---  &	 ---  &    ---  &  II  \\
 161 & 176.560 &   0.154  &  19.77  &  17.33  &  15.59 &     ---  &    ---  &	 ---  &    ---  &  II  \\
 162 & 176.628 &   0.396  &  20.06  &  18.68  &  17.48 &     ---  &    ---  &	 ---  &    ---  &  II  \\
 163 & 176.499 &   0.212  &  20.34  &  18.05  &  16.61 &     ---  &    ---  &	 ---  &    ---  &  II  \\
 164 & 176.496 &   0.194  &    ---  &  17.65  &  15.79 &     ---  &    ---  &	 ---  &    ---  &  II  \\
 165 & 176.520 &   0.186  &  14.97  &  14.69  &  13.45 &     ---  &    ---  &	 ---  &    ---  &  II  \\
 166 & 176.524 &   0.179  &  15.76  &  13.80  &  12.60 &     ---  &    ---  &	 ---  &    ---  &  II  \\
 167 & 176.555 &   0.159  &  11.55  &	8.73  &   7.27 &     ---  &    ---  &	 ---  &    ---  &  II  \\  

\hline          
\end{tabular}
\end{table*}
\end{document}